\documentclass[twocolumn,amssymb,amsmath,
nofootinbib,tightenlines,showpacs,floatfix,
superscriptaddress,aps,prb]{revtex4-1}
\usepackage{amsfonts}
\usepackage{txfonts} 
\usepackage{amsbsy}
\usepackage{bm}%
\usepackage[colorlinks=true,linkcolor=blue,filecolor=blue,menucolor=yellow,urlcolor=blue,citecolor=blue,anchorcolor=blue]{hyperref}
\usepackage{graphicx}
\usepackage{times} 
\usepackage{dcolumn}

\usepackage[justification=justified]{subcaption}
\usepackage{verbatim} 
\begin{document} 

\captionsetup[figure]{justification=raggedright}

\title{Spin current and spin transfer torque in 
ferromagnet/superconductor spin valves} 

\author{Evan Moen}
\email{moenx359@umn.edu}
\author{Oriol T. Valls}
\email{otvalls@umn.edu}
\altaffiliation{Also at Minnesota Supercomputer Institute, University of Minnesota,
Minneapolis, Minnesota 55455}
\affiliation{School of Physics and Astronomy, University of Minnesota, 
116 Church St SE, Minneapolis, Minnesota 55455}

\date{\today}

\begin{abstract} 

Using fully self consistent methods, we 
study spin transport  in realistic, fabricable
experimental spin valve systems consisting of two magnetic
layers, a superconducting layer, and a spacer
normal layer between the ferromagnets. Our methods
ensure that the proper relations between spin current
gradients and spin transfer torques are satisfied.
We present results as a function of geometrical
parameters, interfacial barrier values, misalignment angle between
the ferromagnets,  and bias voltage. 
Our main results are
for the spin current and spin accumulation as  functions of position
within the spin valve structure. 
We see precession of the spin current about the exchange fields
within the ferromagnets, and penetration of the spin current into the superconductor for biases
greater than the critical bias, defined in the text. 
The spin accumulation exhibits oscillating  behavior
in the normal metal, with a strong dependence on the physical parameters 
both as to the structure and formation of the peaks. 
We also study the bias dependence of the 
spatially averaged  spin transfer torque and  spin accumulation. 
We examine the critical bias effect of these quantities, 
 and their dependence on the physical parameters. Our results are
predictive of the outcome 
of future
experiments, as they take into account imperfect interfaces and a realistic geometry.

\end{abstract}


\maketitle

\section{Introduction} 


Spintronic devices, such as spin valves, have seen increasing attention over the years due\cite{tsyzu} to their expected
technological applications (for example,
to non-volatile memory) and for their intrinsic
scientific interest. 
Traditional spin valves\cite{tsyzu} are composed of two ferromagnets ($F$) in close proximity, often separated by a normal metal or insulator. A charge current interacts with the exchange field of the first ferromagnetic component, inducing a polarization in its spin degree of freedom. The second $F$ component is introduced as a spin selector and detector, in which 
a spin current and spin accumulation is predicted and measured\cite{Johnson1993, Jedema2001}. The charge current and the
relative orientation of the exchange fields of the two ferromagnets determine the spin-transport properties of these devices. 
In their application to non-volatile memory, the magnetic memory is current-switched (as opposed to magnetic field-switched) via the spin transfer torque (STT)\cite{berger,Slonczewski1996,Albert2002, myers}. This gives the devices an advantage in power consumption and scalability\cite{Bhatti2017}.

Superconducting spin valves are different. They are spintronic devices that include, in addition to the
$F$ components, one or more layers of a superconducting ($S$) material.
Thus,  superconducting, 
as well as ferromagnetic and normal, components are
involved.  They are
exciting, developing spintronic structures presenting their own unique set of properties and applications\cite{Eschrig2011}. 
In these devices the presence of (usually
traditional, well-understood)
superconductors in proximity to ferromagnetic materials 
fundamentally affects spin transport.
Furthermore, their ultra-low power consumption offers a distinct advantage over standard spin valves, particularly
in memory applications.
Many such devices have been proposed\cite{esch,igor,fomi}. Superconducting spin valves with $F_1/N/F_2/S$ layered structures have been studied\cite{kami,wvhg,zkhv}. The currents in such devices are
in general  spin-polarized and 
can potentially be controlled by STT in nanoscale devices, 
just as in traditional spin valves. However, they are not merely regular spin valves with spin currents. Rather, these are novel 
structures with their own distinct set of spin transport
properties due to the $F/S$ proximity effects\cite{Buzdin2005}. 
Below, we discuss some of
the peculiar properties of these
devices as they are relevant
to our study.

Superconductivity results from the formation of Cooper pairs
consisting of opposite  momentum electrons\cite{BCS}. 
In the usual s-wave superconductivity, these pairs form a singlet state. Ferromagnetism, on the other hand, has a strong tendency to 
break these singlet pairs, while favoring in
principle  
triplet pairing states with $m_z=\pm1$. 
It would seem that ferromagnetism and s-wave superconductivity are largely incompatible. Indeed, the ordinary superconducting 
proximity effects in $F/S$ heterostructures result
in  a heavily damped, oscillatory behavior of the singlet pair
amplitudes in the $F$ layer regions\cite{Buzdin1990, Halterman2002}, caused by Cooper pairs acquiring a center of mass momentum\cite{demler}. This oscillatory behavior is critical to understanding $F/S$ heterostructures, 
as it makes all transport measurements highly dependent on the thicknesses of each material layer. 
However, proximity effects in $F/S$ 
structures are by no means limited to  those arising
from the s-wave Cooper pairs in the $S$ material. Indeed, there are
long range proximity effects from 
triplet pair correlations that are induced in the structure
by the presence of nonuniform  exchange fields\cite{berg86,Halterman2009,hbv, bvh,zdravkov2013}.  This
conversion is possible because, unless all exchange fields
are collinear, the Hamiltonian 
does not commute with $S_z$, the $z$ component of the Cooper pair spins:
 thus it 
is not conserved. 

Because of the Pauli principle, the triplet correlations must be odd in frequency\cite{berezinskii} or equivalently in time\cite{bvh}. 
In the presence of a uniform exchange field, only the $m_z=0$ triplet component may be induced. The required
non-uniform exchange field can be introduced
in a variety of ways: for example
one can have a $F_1/F_2/S$ heterostructure with noncollinear
exchange fields, or 
a single $F$ layer with a non-uniform magnetization texture such as
one may have with magnetic domains or, in a more controllable 
way, by using a magnet
such as  Holmium\cite{chiodi,hoprl,ho,gu2015} in which the
magnetic structure is spiral.
In these cases the presence of $m_z=\pm1$ pairs
is compatible with conservation laws and the Pauli
principle, and in fact such 
pairs  are 
usually induced. The exchange fields do not 
necessarily break these triplet correlations, and thus 
the proximity effect can be  long ranged\cite{bvermp,Eschrig2008,leksin,Bergeret2007,kalcheim,singh,ha2016} in  $F$.  In heterostructures which include two ferromagnetic
layers $F_1$ and $F_2$,  as we consider
in this paper, one can immediately see that 
there will be an interesting angular dependence 
of the results on the misalignment angle $\phi$ between the two $F$
layers, as their orientations vary from being
parallel, to orthogonal, to antiparallel. 
In traditional spin valves, this angular dependence 
is characterized by the magnetoresistance
obtained by comparing the parallel (P) and antiparallel (AP) configurations\cite{fert}. In the
superconducting devices, as triplet  pairs are induced, 
singlet pair amplitudes decrease, diminishing
 the strength of the superconducting pair potential and influencing the transport properties\cite{Moen2017,wvhg}. As $\phi$
 is varied between $0^\circ$ and $180^\circ$  a unique angular dependence that is nonmonotonic is produced.

The superconducting proximity effects discussed above affect
both the thermodynamic and the transport
properties of  the device. A fundamental
contribution to both arises from Andreev reflection\cite{Andreev}
at the interfaces. 
Andreev reflection is the process of electron-to-hole conversion by the creation or annihilation of a Cooper pair, occurring at the interface of a superconductor. There are two types of Andreev reflection: conventional and anomalous. In conventional Andreev reflection, 
the reflected electron/hole has spin opposite  to that of the incident particle. 
In anomalous Andreev reflection, these electron/hole pairs have the same spin. It has  been shown\cite{wvhg,linder2009,visani,niu,ji} that 
normal and anomalous Andreev reflection are correlated with 
triplet proximity effects. Understanding and accurately characterizing the transmission amplitudes of the Andreev reflections is pertinent to all transport calculations in superconducting  heterostructures\cite{btk,tanaka,beenakker,zv2}, particularly for quantities with spatial dependence such as the spin current and spin transfer torque. 

The practical fabrication of $F/F/S$ valve structures results in devices 
that deviate very significantly from theoretical idealizations. To
be able to modify the angle $\phi$ requires the insertion
of a normal metal spacer between the $F$ layers, so that
they are decoupled and the magnetization
of one of them can be rotated
individually. In addition,
even high quality interfaces between all layers involved are not perfect:
some interfacial scattering is
inevitable and transport\cite{igor} in superconducting
spin valves is very
sensitive to it\cite{zv2,Moen2017}, as is also
the case\cite{Takahashi2003} for 
spin transport in traditional spin valves. It has been
shown that, if the the normal spacer and the
interfacial scattering are properly taken into account,
then it is
possible to quantitatively characterize to high
accuracy\cite{alejandro} the thermodynamic properties
of such devices. In recent work\cite{Moen2017}, 
we have also examined the charge transport properties of $F_1/N/F_2/S$ heterostructures with an emphasis on practical, realistic layer 
thicknesses and interfacial scattering parameters.
However, spin transport properties, such as spin-current and 
the STT were calculated only for the ``proof
of principle''  ideal case with no normal metal spacer or 
interfacial scattering parameters.
 
In this paper, we perform spin transport
calculations for  fabricable
samples. We assume  realistic geometrical parameters (thickness
of the layers, including that of $N$) and 
material parameters appropriate 
to the Co and Nb layers used in experiments\cite{alejandro}. 
The charge and spin transport properties depend strongly on 
the applied bias voltage. 
Many of their features\cite{wvhg, Moen2017}  change
rather abruptly when the applied voltage
reaches  the critical bias (CB) value,
which is related to  the self-consistent pair potential 
within the  superconductor. This value
is less than the pair potential  bulk value due to the proximity effects. 
The transport properties are
quite different for an applied 
voltage bias below and above the CB. 
This effect is also dependent on the misalignment angle of the 
exchange fields, usually in a  nonmonotonic\cite{Moen2017}
way. 
Here, we examine the dependence of the spin-transport properties 
on the layer thicknesses, the importance of which has been mentioned above, the interfacial scattering strengths,
and the applied bias voltage, including CB effects. 
We hope to establish a broad understanding of how sample quality 
and geometry affect spin transport results in $F_1/N/F_2/S$ systems so that they may then be compared to experimental results. 

In our calculations, we use a self consistent solution to the 
Bogoliubov de Gennes (BdG) equations\cite{degennes} to 
calculate the pair potential, and then
employ this potential in the transport calculations
via a transfer matrix method\cite{Moen2017}. This method correctly
incorporates the normal and
Andreev reflection and transmission amplitudes of the electrons and holes. We evaluate then 
the spin current, the STT, and the magnetization, all as  functions
of position within the $F_1/N/F_2/S$ heterostructure and  of the applied 
bias. We examine their dependence on the misalignment angle $\phi$. 
We also vary the layer thickness, within realistic
limits, and the interfacial scattering strengths.
Our focus will be the analysis of the physical parameters for experimental use, as well as on the underlying physics of the spin transport. 

Spin transport is considerably more complex than charge transport.
As opposed to the charge current, which is a constant
through the sample due to charge conservation, the
spin current varies with position, and this variation
is related to the STT. Furthermore, since spin is a vector
the spin current is in principle a tensor, although it does
reduce to a vector in spin space
in the quasi-one dimensional geometry we will consider here. 
Thus all quantities are spatially dependent. Together
with  the spatially
oscillatory nature of the singlet and triplet amplitudes, we find a strong and intricate dependence of spin transport
on the layer thicknesses. Furthermore, the proximity effects are particularly influential on the spin transport properties,  as they relate to 
the spin-pairing and the induced triplets. 
We thus see a nonmonotonic dependence on 
$\phi$, as well as a strong dependence on the interfacial scattering strengths. Interfacial scattering  generally inhibits the proximity effects but, because 
there are several barriers,
resonance features such as those found for charge transport\cite{Moen2017}
can also arise. We will also analyze the 
average of the spin transport 
quantities over each layer: we have
found this particularly useful in
studying the bias dependencies and in better establishing
the underlying physical principles at work. We hope through this work
 to provide future experiments with some deeper context as to how these parameters may affect their results.

After this Introduction, we briefly review our methods for transport calculations in Sec. \ref{methods}. The results, as well as their discussion, are presented in Sec. \ref{results}. We summarize our work in Sec. \ref{conclusions}.

\section{Methods}
\label{methods}

\subsection{The basic equations}

\label{static}

\begin{figure}
\includegraphics[width=0.45\textwidth] {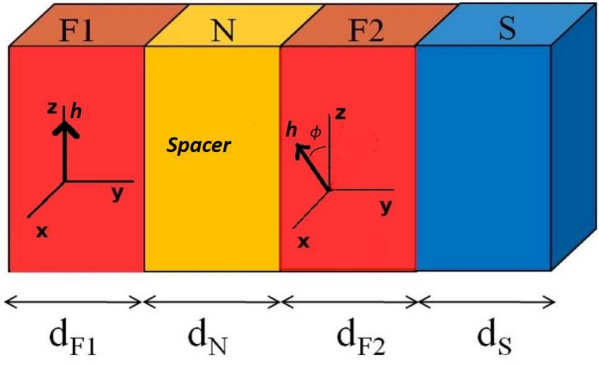} 
\caption{Scheme of the system studied.  The exchange field of 
the second ferromagnet $F_2$ is rotated in the $x-z$ plane by an angle $\phi$.
The direction of the transport is in the $y$ direction. The thicknesses are not to scale (see text).} 
\label{figure1}
\end{figure}

The geometry of the system we study is depicted in Fig.~\ref{figure1}. 
The layers are assumed infinite in the transverse, $x-z$ plane, and
have  finite widths in the $y$ direction. This assumption makes the system quasi-one-dimensional. The magnetizations of the outer ($F_1$) and inner ($F_2$) layers are misaligned
by an angle $\phi$ in the $x-z$ plane.  Below, we 
briefly summarize our methods and procedures which 
are ultimately based in Ref.~\onlinecite{degennes} 
and are described extensively in Refs.~\onlinecite{wvhg,Moen2017}.

The Hamiltonian appropriate to our system is
\begin{eqnarray}
\label{ham}
{\cal H}_{eff}&=&\int d^3r \left\{ \sum_{\alpha}
\hat{\psi}_{\alpha}^{\dagger}\left(\mathbf{r}\right){\cal H}_0
\hat{\psi}_{\alpha}\left(\mathbf{r}\right)\right.\nonumber \\
&+&\left.\frac{1}{2}\left[\sum_{\alpha,\:\beta}\left(i\sigma_y\right)_{\alpha\beta}
\Delta\left(\mathbf{r}\right)\hat{\psi}_{\alpha}^{\dagger}
\left(\mathbf{r}\right)\hat{\psi}_{\beta}^{\dagger}
\left(\mathbf{r}\right)+H.c.\right]\right.\nonumber \\
&-&\left.\sum_{\alpha,\:\beta}\hat{\psi}_{\alpha}^{\dagger}
\left(\mathbf{r}\right)\left(\mathbf{h}\cdot\bm{\sigma}
\right)_{\alpha\beta}\hat{\psi}_{\beta}\left(\mathbf{r}\right)\right\},
\end{eqnarray}
where $\Delta\left(\mathbf{r}\right)$ is the pair potential, 
and $\mathbf{h}$ is the Stoner field. The field $\mathbf{h}$ is taken along the $z$ axis in the outer ferromagnetic layer $F_1$ and forms an angle $\phi$ with the $z$ axis in the inner ferromagnetic layer $F_2$. This field is then zero in the superconductor $S$ and normal metal spacer $N$. We have assumed equal magnitude of the fields $h_1=h_2 \equiv h$ since in
experiments the same material is 
typically  employed for both ferromagnetic layers. 
${\cal H}_0$ is the single-particle Hamiltonian, and
it includes the interfacial scattering. The indices $\alpha$ and $\beta$ denote spin indices and $\sigma_i$ are the Pauli matrices.

Performing a generalized Bogoliubov transformation, we take $\psi_{\sigma}=\sum_n\left(u_{n\sigma}\gamma_n 
+\eta_{\sigma}v_{n\sigma}^{\ast}\gamma_n^{\dagger}\right)$
where $\eta_\sigma \equiv 1(-1)$ for spin down (up), and $u_{n \sigma}(\mathbf{r})$ and $v_{n \sigma}(\mathbf{r})$ are the spin-dependent quasiparticle and quasihole amplitudes. Due to the geometry of the system being quasi-one-dimensional, the spatial dependence 
on $\mathbf{r}$ becomes a dependence on  $y$ alone. Then, we can rewrite the eigenvalue equation corresponding to the Hamiltonian given by Eq. (\ref{ham}) as

\begin{align}
&\begin{pmatrix}
{H}_0 -h_z&-h_x&0&\Delta(y) \\
-h_x&{H}_0 +h_z&\Delta(y)&0 \\
0&\Delta(y)&-({H}_0 -h_z)&-h_x \\
\Delta(y)&0&-h_x&-({H}_0+h_z) \\
\end{pmatrix}
\begin{pmatrix}
u_{n\uparrow}(y)\\u_{n\downarrow}(y)\\v_{n\uparrow}(y)\\v_{n\downarrow}(y)
\end{pmatrix} \nonumber \\
&=\epsilon_n
\begin{pmatrix}
u_{n\uparrow}(y)\\u_{n\downarrow}(y)\\v_{n\uparrow}(y)\\v_{n\downarrow}(y)
\end{pmatrix}\label{bogo},
\end{align}

We use natural units $\hbar=k_B=1$. The quasi-one-dimensional Hamiltonian 
is $H_0=-(1/2m)(d^2/dy^2)+\epsilon_\perp-E_F(y)+U(y)$ where $\epsilon_\perp$ is the transverse energy, so that Eq.~(\ref{bogo}) is a set of decoupled equations, one
for each $\epsilon_\perp$. The energy 
bandwidth $E_F$ can in principle
be  layer dependent. In the $S$ layer, for
example, we write  $E_F(y)=E_{F S} \equiv k_{FS}^2/{2m}$. 
$U(y)$ is the interfacial scattering, which we take to be spin independent in the form $U(y)=H_1\delta(y-d_{F1})+H_2\delta(y-d_{F1}-d_N)+H_3\delta(y- d_{F1}-d_N-d_{F2})$ where $H_i$ are the scattering strengths of the respective interfaces. These scattering strengths are best characterized by the dimensionless parameters $H_{B i} \equiv H_i/v_F$, where $v_F$ is the Fermi speed in $S$. These scattering parameters are quite
essential to characterizing possible devices, as even for clean interfaces, some scattering due to residual surface roughness is inevitable.
Transport results turn out to be much more sensitive than
thermodynamic quantities to interfacial scattering.

All of the  calculations must be done self-consistently to preserve charge conservation\cite{Moen2017,wvhg}. The self-consistency 
condition allows for the proper
inclusion of the proximity effect, which is of primary importance
to our study. The self consistency condition is:
\begin{equation}
\label{del}
\Delta(y) = \frac{g(y)}{2}{\sum_n}^\prime
\bigl[u_{n\uparrow}(y)v_{n\downarrow}^{\ast}(y)+
u_{n\downarrow}(y)v_{n\uparrow}^{\ast}(y)\bigr]\tanh\left(\frac{\epsilon_n}{2T}\right), \,
\end{equation}
where $g(y)$ is the superconducting coupling constant in the singlet channel 
and it is nonzero in the $S$ layer only. The sum is over  eigenvalues,
and the prime symbol indicates
that the sum is limited to states with eigenenergies within a cutoff $\omega_D$ from 
the Fermi level. The self-consistency procedure is this: 
we start with a suitable choice for $\Delta(y)$, compute the quasi-particle 
and quasi-hole amplitudes using Eq.~(\ref{bogo}), and obtain $\Delta(y)$ using Eq.~(\ref{del}). Then we repeat this process, substituting 
the iterated $\Delta(y)$ until the input of Eq.~(\ref{bogo}) matches the output of Eq.~(\ref{del}). 
Self-consistency is fundamental in all transport calculations. 
It is a prerequisite for charge conservation\cite{wvhg,bagwell,sols2,sols}. 
From the Heisenberg equation we have:

\begin{equation}
\label{cons}
\frac{\partial}{\partial t}\left\langle\rho({\mathbf r})\right\rangle
=i\left\langle\left[{\cal H}_{eff},\rho({\mathbf r})\right]\right\rangle.
\end{equation}
where $\rho({\mathbf r})$ is the charge density.
In the steady-state, and in our geometry, we can rewrite  this as:

\begin{equation}
\frac{\partial j_y(y)}{\partial y}= 2e {\rm Im}\left\{\Delta(y)\sum_n\left[u_{n \uparrow}^* 
v_{n \downarrow}+u_{n\downarrow}^*v_{n\uparrow}\right]\tanh\left(\frac{\epsilon_n}{2T}\right)\right\}.  
\label{currentuv}
\end{equation}
Charge conservation is preserved if ${\partial j_y(y)}/{\partial y}$ is identically zero, which is  guaranteed when the self-consistency 
condition Eq.~(\ref{del}) is applied. Another
 reason why transport is dependent on self-consistency is more obvious: as the pair potential changes, so does the energy spectrum within the superconductor. Proper inclusion of ordinary
and Andreev reflection at the interfaces is obviously necessary
for a proper account of the transport properties of heterostructures, and the variation of the self-consistent pair amplitudes is most pronounced at the superconducting interface due to proximity effects. Therefore, it is 
mandatory that we calculate transport using a fully self-consistent pair potential. 

\subsection{Spin transport Quantities}
\label{spintran} 

The spin transport related quantities we consider
are the spin current, the STT, and the local magnetization. These
are all studied as functions of applied
bias voltage $V$. We aim to describe the  position 
dependence of these bias-dependent quantities
within the multi-layer structure, for a range of
relevant values of the geometrical parameters, including
$\phi$. 
 In our geometry
the spin current is a vector in spin space:
\begin{equation}
\label{spincurrents}
S_i\equiv\frac{i\mu_B}{2m}\sum_\sigma\left\langle \psi_\sigma^{\dagger}\sigma_i\frac{\partial \psi_\sigma}{\partial y}
-\frac{\partial \psi_\sigma^{\dagger}}{\partial
y}\sigma_i\psi_\sigma\right\rangle.
\end{equation}
The spin current density is not a conserved quantity within the ferromagnetic regions. We can relate its gradient to the
local magnetization  ${\bf m}\equiv -\mu_B\sum_{\sigma}\psi_\sigma^\dagger{\bf \sigma}\psi_\sigma$, where $\mu_B$ is the Bohr magneton,
 by writing the continuity equation for the local magnetization 
in the form:
\begin{equation}
\label{spinconserve}
\frac{\partial}{\partial t}\langle m_i \rangle+ \frac{\partial}{\partial y} S_i= \tau_i,
\enspace i=x,y,z 
\end{equation}
where $\bm{\tau}$ is the spin-transfer torque $\bm{\tau}\equiv 2\mathbf{m}\times\mathbf{h}$. In the steady state, 
 ${\partial} m_i / \partial t$ is zero. This means 
 that the spin current will not be constant within the ferromagnetic layers, and that the local magnetization, even in the steady state, 
 is intrinsically tied to the spin current via the STT. 
 
  We can write the magnetization 
 and the spin current in terms of the 
 self consistent quasi particle and quasi hole amplitudes. In the low temperature limit, the 
 expression for the local magnetization reads\cite{wvhg},
\begin{widetext}
\begin{subequations}
\label{mag} 
\begin{align}
m_x=-\mu_B\left[\sum_n\left(-v_{n\uparrow}v_{n\downarrow}^{\ast}-v_{n\downarrow}v_{n\uparrow}^{\ast}\right)\right. 
\left.+\sum_{\epsilon_\mathbf{k}<eV}\left(u_{\mathbf{k}\uparrow}^{\ast}u_{\mathbf{k}\downarrow}
+v_{\mathbf{k}\uparrow}v_{\mathbf{k}\downarrow}^{\ast}
+u_{\mathbf{k}\downarrow}^{\ast}u_{\mathbf{k}\uparrow}
+v_{\mathbf{k}\downarrow}v_{\mathbf{k}\uparrow}^{\ast}\right)\right]\\
m_y=-\mu_B\left[i\sum_n\left(v_{n\uparrow}v_{n\downarrow}^{\ast}-v_{n\downarrow}v_{n\uparrow}^{\ast}\right)\right. 
\left.-i\sum_{\epsilon_\mathbf{k}<eV}\left(u_{\mathbf{k}\uparrow}^{\ast}u_{\mathbf{k}\downarrow}
+v_{\mathbf{k}\uparrow}v_{\mathbf{k}\downarrow}^{\ast}
-u_{\mathbf{k}\downarrow}^{\ast}u_{\mathbf{k}\uparrow}
-v_{\mathbf{k}\downarrow}v_{\mathbf{k}\uparrow}^{\ast}\right)\right]\\
m_z=-\mu_B\left[\sum_n\left(|v_{n\uparrow}|^2-|v_{n\downarrow}|^2\right)\right.
\left.+\sum_{\epsilon_\mathbf{k}<eV}\left(|u_{\mathbf{k}\uparrow}|^2
-|v_{\mathbf{k}\uparrow}|^2
-|u_{\mathbf{k}\downarrow}|^2
+|v_{\mathbf{k}\downarrow}|^2\right)\right],
\end{align}
\end{subequations}
where the first terms on the right side are
 the ground state local magnetization components,
and the second terms denote the bias dependent contributions.
We can  define  a direct analog of the spin accumulation by 
removing the first terms on the right
side   
 $\delta {\bf m}(V)\equiv {\bf m}(V)-{\bf m}(0)$, 
revealing the change in magnetization due to the finite bias.

We can use the same procedure for the spin current components, Eq.~(\ref{spincurrents}),
and expand in terms of the $u_n$ and $v_n$ wavefunctions. In the $T=0$ limit the result is\cite{wvhg}:
\begin{subequations}
\label{spincur}
\begin{align}
S_x=\frac{-\mu_B}{m}{\rm Im}\left[\sum_n\left
(-v_{n\uparrow}\frac{\partial v_{n\downarrow}^{\ast}}{\partial y}
-v_{n\downarrow}\frac{\partial v_{n\uparrow}^{\ast}}{\partial y}\right)\right.
\left.+\sum_{\epsilon_\mathbf{k}<eV}\left(u_{\mathbf{k}\uparrow}^{\ast}\frac{\partial u_{\mathbf{k}\downarrow}}{\partial y}
+v_{\mathbf{k}\uparrow}\frac{\partial v_{\mathbf{k}\downarrow}^{\ast}}{\partial y}
+u_{\mathbf{k}\downarrow}^{\ast}\frac{\partial u_{\mathbf{k}\uparrow}}{\partial y}
+v_{\mathbf{k}\downarrow}\frac{\partial v_{\mathbf{k}\uparrow}^{\ast}}{\partial y}\right)\right]\\
S_y=\frac{\mu_B}{m}{\rm Re}\left[\sum_n\left(-v_{n\uparrow}
\frac{\partial v_{n\downarrow}^{\ast}}{\partial y}
+v_{n\downarrow}\frac{\partial v_{n\uparrow}^{\ast}}{\partial y}\right)\right.
\left.+\sum_{\epsilon_\mathbf{k}<eV}\left(u_{\mathbf{k}\uparrow}^{\ast}\frac{\partial u_{\mathbf{k}\downarrow}}{\partial y}
+v_{\mathbf{k}\uparrow}\frac{\partial v_{\mathbf{k}\downarrow}^{\ast}}{\partial y}
-u_{\mathbf{k}\downarrow}^{\ast}\frac{\partial u_{\mathbf{k}\uparrow}}{\partial y}
-v_{\mathbf{k}\downarrow}\frac{\partial v_{\mathbf{k}\uparrow}^{\ast}}{\partial y}\right)\right]\\
S_z=\frac{-\mu_B}{m}{\rm Im}\left[\sum_n\left(v_{n\uparrow}\frac{\partial v_{n\uparrow}^{\ast}}{\partial y}
-v_{n\downarrow}\frac{\partial v_{n\downarrow}^{\ast}}{\partial y}\right)\right.
\left.+\sum_{\epsilon_\mathbf{k}<eV}\left(u_{\mathbf{k}\uparrow}^{\ast}\frac{\partial u_{\mathbf{k}\uparrow}}{\partial y}
-v_{\mathbf{k}\uparrow}\frac{\partial v_{\mathbf{k}\uparrow}^{\ast}}{\partial y}
-u_{\mathbf{k}\downarrow}^{\ast}\frac{\partial u_{\mathbf{k}\downarrow}}{\partial y}
+v_{\mathbf{k}\downarrow}\frac{\partial v_{\mathbf{k}\downarrow}^{\ast}}{\partial y}\right)\right], 
\end{align}
\end{subequations}
\end{widetext}
where again the first terms on the right side
are the spin current density at zero bias, and the second terms the contribution from the applied bias. This calculation is independent of that of the local magnetization. Thus we can verify the relation between the STT and the spin current in Eq.~(\ref{spinconserve}), as has previously been pointed out\cite{wvhg,Moen2017}.

\subsection{Transfer Matrix Method and Spin Transport}
\label{transport}

Here, we give a brief summary of our spin transport calculation 
methodology. An extensive explanation has been given in 
Ref.~\onlinecite{wvhg}. We review these methods primarily because 
Ref.~\onlinecite{wvhg} focused on charge transport, and
 it is useful to clarify how they extend to spin transport, 
which requires some extra care. 

The procedure to calculate
the conductance $G(V)$ involved  merely evaluating the reflection and transmission amplitudes governed by the continuity of the wavefunction 
and discontinuity of its derivatives. This has to be done 
at each interface 
for both particles and holes, and for each spin, i.e.
including both ordinary and Andreev
reflection, as one 
would do in elementary quantum mechanics. In the $S$ electrode, the
procedure
is\cite{wvhg} to divide it into arbitrarily thin layers,
in each of which the $y$-dependent self-consistent pair potential, as 
previously determined numerically, can be replaced by a constant. 

In the expressions for the local magnetization Eqn.~(\ref{mag}) and the spin
current Eqn.~(\ref{spincur}) we have two terms in the right sides.
The first is the equilibrium result, and can be calculated straightforwardly
by the methods of Section \ref{static}. The more important
terms are, of course, the bias driven contributions. To evaluate 
those we have to rebuild the wavefunctions so that they correspond
to the proper boundary conditions of injected spin up or spin down
particles (see e.g. Eqns.~(4) and (5) of Ref.~\onlinecite{wvhg} or 
Ref.~\onlinecite{Moen2017}). The method is
in essence nothing but the elementary quantum mechanical procedure
of building plane wave solutions out of stationary state
wavefunctions, but it is mathematically
much more complicated. The procedure is as fully described in
Ref.~\onlinecite{wvhg}    
except for the 
presence of the $N$ layer, which can be included by
 a trivial extension of either an $F$ layer 
with $h$ taken to be zero, or an $S$ layer with $\Delta=0$. 
The transfer matrix method simply transcribes the continuity conditions for 
each amplitude, and the discontinuity in the derivatives
arising from the delta function interfacial
scattering, to each adjacent layer. From these rebuilt wavefuctions the
second terms in the right sides of the expressions for ${\bf m}(y)$
and ${\bf S}(y)$ are  straightforwardly calculated by adding the
appropriate contributions. 
This procedure is especially important in spin transport calculations, 
as the quantities involved
depend on position and the simple
BTK\cite{btk} procedure that one  employs
for the conductance does not apply.

\section{Results}
\label{results}
\subsection{General}

We report on the spin transport quantities, specifically
the spin current, the spin transfer torque, and the bias-dependent portion of the magnetization, which as mentioned above is a measure
of the spin accumulation. Each of these quantities
depends on the applied bias voltage $V$,
which we normalize to $E\equiv eV/{\Delta_0}$, where
$\Delta_0$ is the bulk value of the pair potential
in bulk $S$ material. These quantities
depend also on the position $y$ within the
sample.  All lengths are normalized by
$k_{FS}$, and normalized lengths are denoted by
the corresponding capital letter, e.g. $Y \equiv k_{FS}y$. All energies except for the bias are normalized to the Fermi energy in $S$. The magnetization 
components $m_i$ are 
normalized by $-\mu_B (N_\uparrow + N_\downarrow)$, and, correspondinly,
the spin current $S_i$ is normalized\cite{wvhg} by 
 $-\mu_B (N_\uparrow + N_\downarrow)E_{FS}/k_{FS}$. 
The normalization of the scattering strength parameters has been 
introduced above: values in excess of unity correspond
to a tunneling limit situation. We will assume that the two ferromagnetic 
materials are the same, and hence take
the field strengths $h_1$ and $h_2=h$ to be equal. We will use
the value $h=0.145$ in our dimensionless
units. This value was shown to be appropriate
to describe the transition temperature\cite{alejandro} of similar
samples in which Co was the ferromagnetic material. Similarly, we will
assume that the 
scattering strengths for the 
two $N/F$ interfaces are the same $H_{B1}=H_{B2}\equiv H_B$. 
We will take the 
effective coherence length of the superconducting order parameter to be $\Xi_0=115$ which was found to be appropriate for 
samples in which the $S$ layer was Niobium\cite{alejandro}. We set the superconducting layer thickness to be $D_S=180$, which is large enough compared to $\Xi_0$ to allow for superconductivity, but not so large that the proximity effect is negligible within the superconductor. This has been shown in previous results\cite{Moen2017} to provide a more prominent critical bias feature in charge transport due to the variation in the pair potential $\Delta(y)$. 
We will also fix the thickness of the outer ferromagnet to
$D_{F1}=30$ as we have found that the results are less sensitive
to this parameter. We will  consider variations of $D_N$ and $D_{F2}$.
We have  assumed that any band mismatch
parameters are unity. Although this
is not generally true in real systems, in practice
the effects of such a mismatch can be incorporated into the 
effective value
of the scattering strength parameter when interpreting
and fitting data. 

Below, we will be showing results for six different 
sets of the parameters $D_{F2}, D_N, H_B, H_{B3}$.
 For each set of parameters we will examine
the following vector 
quantities: the spin current,  the spin accumulation,  the 
spatially averaged spin accumulation in $S$ and $N$, 
and  the spatially averaged STT in both $F$ layers. 
For the first two, we 
will examine each component at low-bias, $E=0.6$, and 
at high-bias values, $E=2$. We will study the quantities 
$\delta m_i\equiv m_i(V)-m_i(0)$ 
and $\tau_i$ as a function of the bias, rather
than of position, by averaging these quantities
over a layer. Thus, for example
 $\langle \tau_i\rangle \equiv 1/D_\ell \int dY \tau_i(Y)$ where 
the integral is over the relevant layer, of thickness $D_\ell$.
In all cases we plot the results
for several values of the angular mismatch angle $\phi$.
The number of quantities involved for each set of physical parameters is excessively large, therefore
 we focus on only the most remarkable features and angular dependencies, 
and on their distinctive
behavior as a function of the physical parameters.

\subsection{Ideal Interfaces}
\begin{figure*} 
  \centering
  \vspace{-1cm}
  \begin{subfigure}[b]{.48\textwidth}
    \hspace*{-1.7cm}\includegraphics[angle=-90,width=1.3\textwidth] {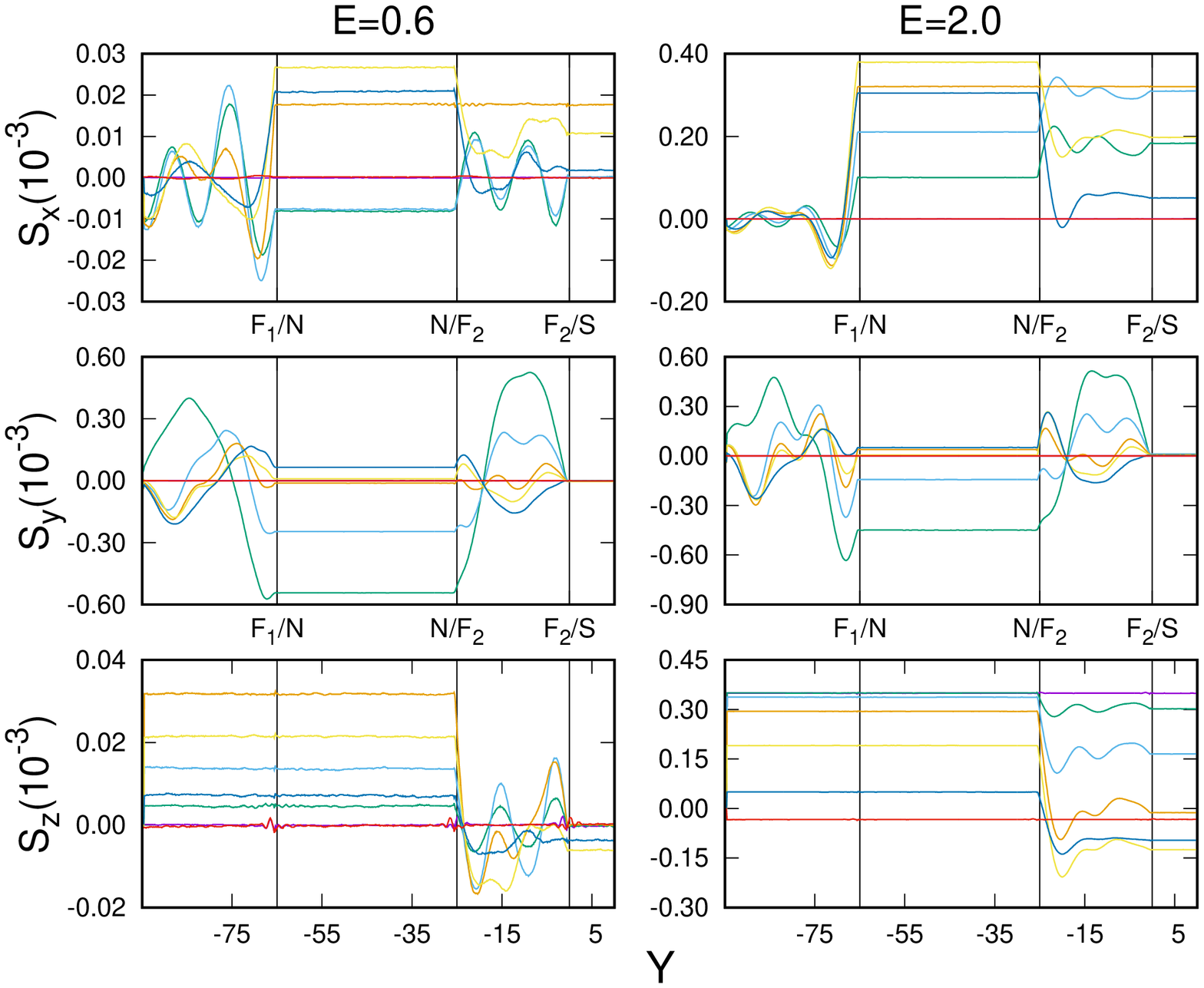} 
    \vspace*{-0.1cm}\small\caption{Local Spin Current}
    \label{Fig2a}
  \end{subfigure}
  \hfill
  \begin{subfigure}[b]{.48\textwidth}
    \hspace*{-1.1cm}\includegraphics[angle=-90,width=1.3\textwidth] {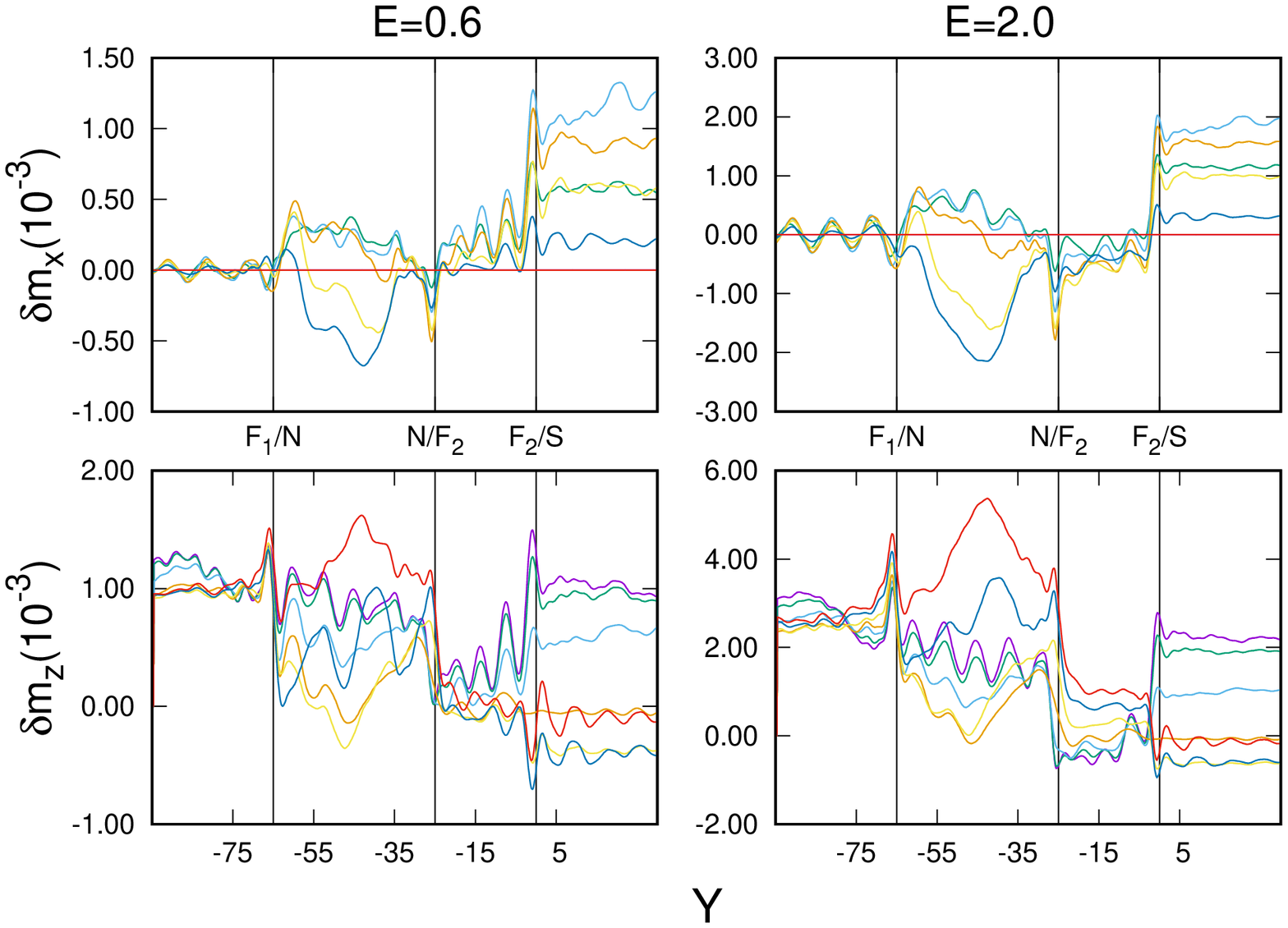}\vspace*{-0.6cm}
    \small\caption{Local Spin Accumulation}
    \label{Fig2b}
  \end{subfigure}
  \vskip\baselineskip
  \vspace*{-1.7cm}
  \begin{subfigure}[b]{.48\textwidth}
    \hspace*{-1.3cm}\includegraphics[angle=-90,width=1.3\textwidth] {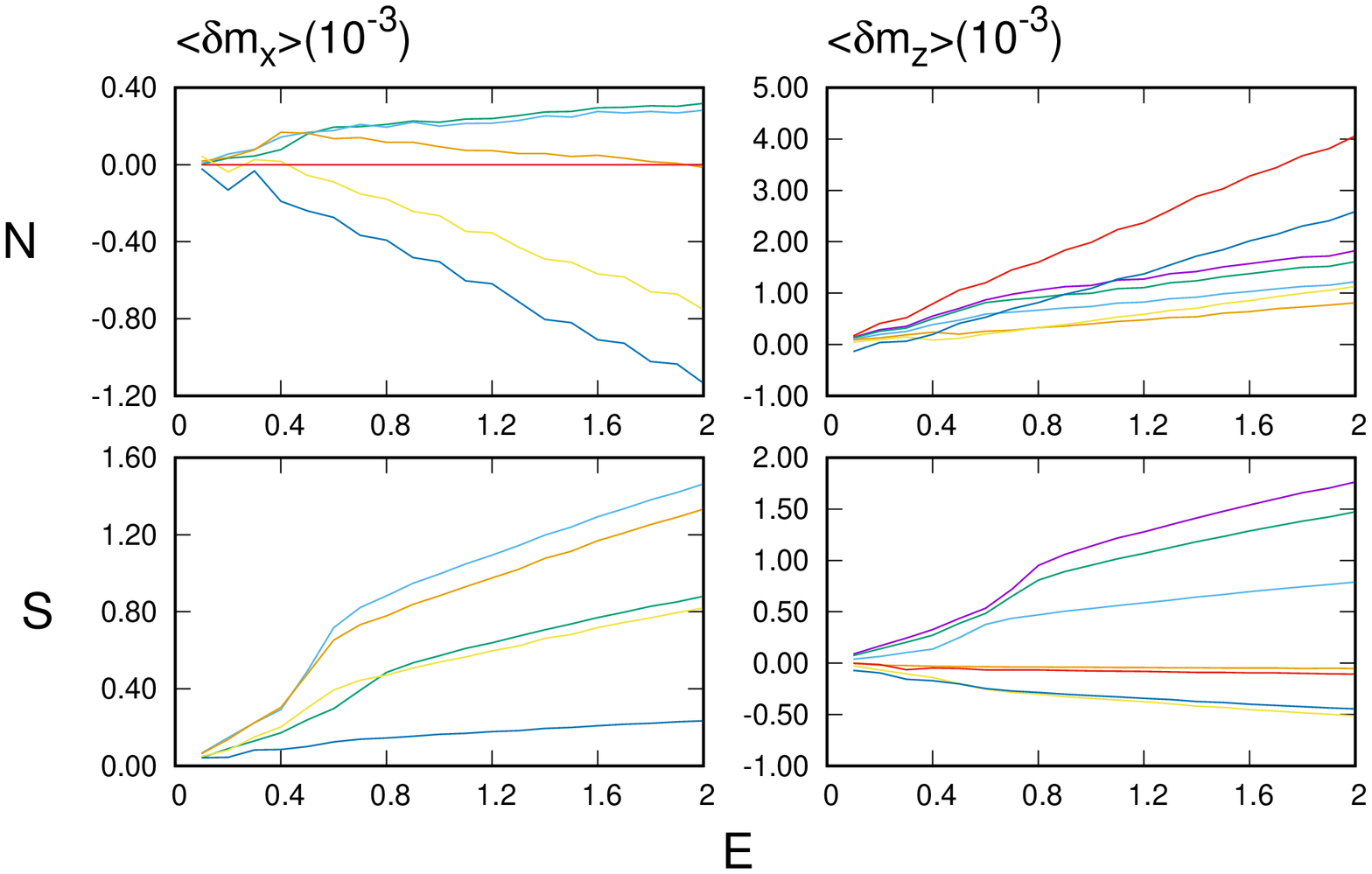} 
    \vspace*{-1.0cm}\small\caption{Spatially Averaged Local Spin Accumulation}
    \label{Fig2c}
  \end{subfigure}
  \quad
  \begin{subfigure}[b]{.48\textwidth}
    \hspace*{-1.7cm}\includegraphics[angle=-90,width=1.3\textwidth] {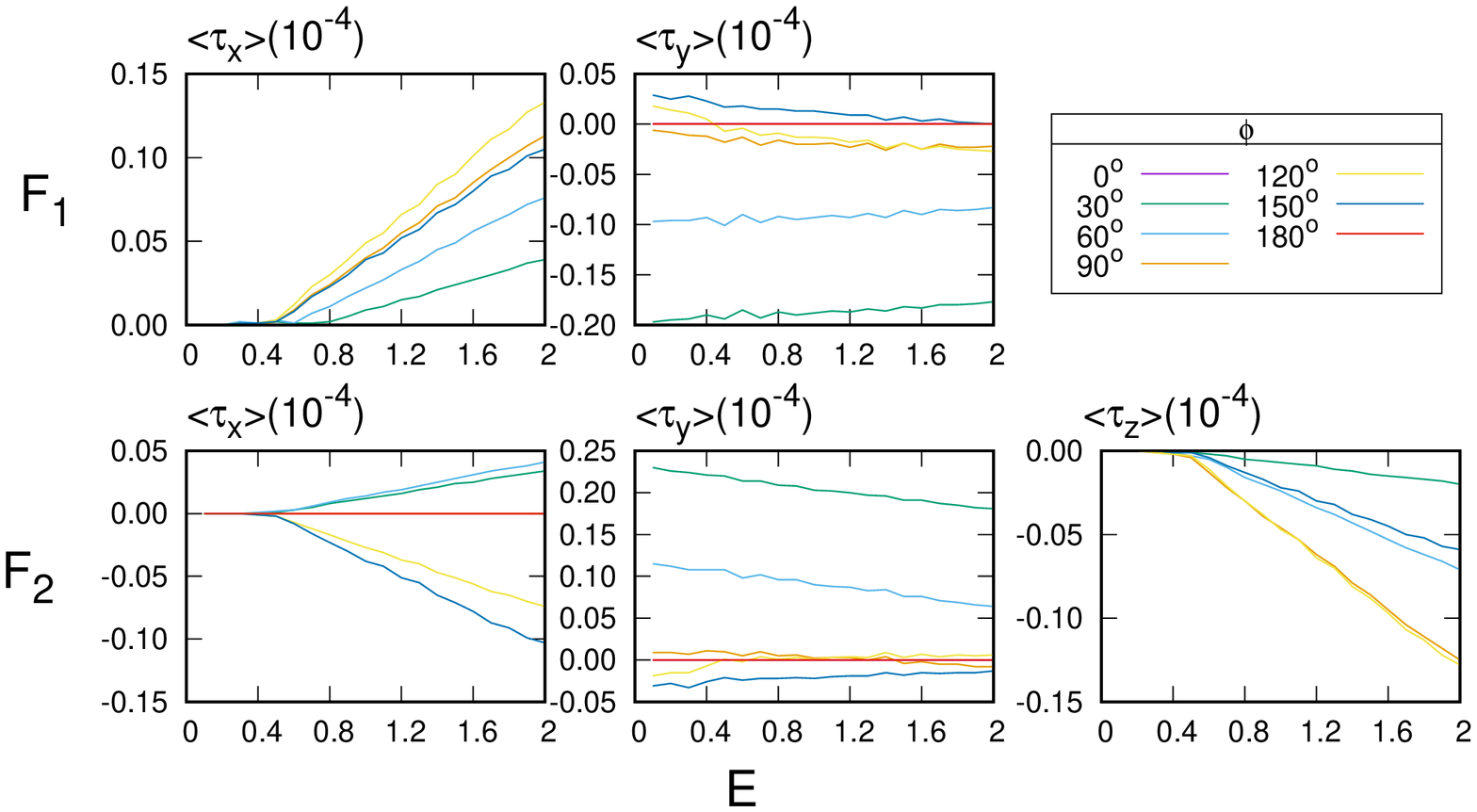}\vspace*{-1.7cm}
    \small\caption{Spatially Averaged Spin Torque}
    \label{Fig2d}
  \end{subfigure}
\caption{Results for ideal interfaces.  
The layer thicknesses for the $F_1/N/F_2/S$ layers are
$30/40/25/180$ respectively, and the interfacial barriers $H_B$ and $H_{B3}$ are both zero. The key for the angular dependence is 
in the upper right panel of set (d). See text for details.}
\label{figure2}
\end{figure*}

In Fig.~\ref{figure2} we show the results for
a physical parameter set with ideal
interfaces (zero interfacial scattering). The layer thicknesses 
for the $F_1/N/F_2/S$ layers are
$30/40/25/180$ respectively. 
This case can be
compared with  previous results\cite{Moen2017} obtained
in some particular cases 
in the absence of
the normal metal layer $N$. 
The normal layer greatly reduces the STT at the interfaces between the ferromagnets. 
We start by examining the fundamental features of each quantity mentioned, 
as a baseline for comparison with subsequent figures. The set of 
panels labeled (a) show the components of the spin current
as a function of position, and the set labeled (b) the
spin accumulation, also as a function
of position.  Sets (c) and (d) refer to 
the spatially averaged spin
accumulation and STT respectively, as functions of bias.

\begin{figure}
 \hspace*{-0.7cm}\includegraphics[width=0.50\textwidth] {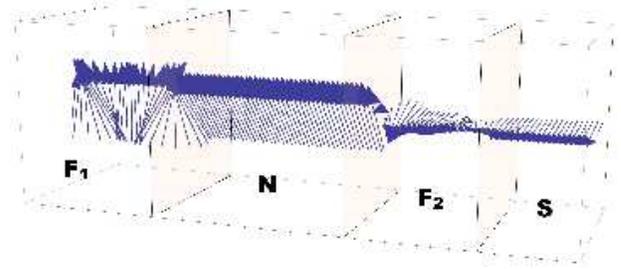}
\caption{A 3D representation of the spin current from Fig. \ref{Fig2a} at $\phi=90^\circ$ and $E=2.0$. From left to right, the boxes comprise the layers
$F_1/N/F_2/S$ respectively. The spin current precesses about the exchange field in $F$, while also dampening in $F_2$. The orientation of the field in $S$ is rotated to $90^\circ$ from the $z$ axis.}
\label{figure3D}
\end{figure}

In Fig.~\ref{Fig2a}, we examine the spin current components $S_i$ 
(top to bottom) as a function of position $Y$
at low to high bias 
($E=0.6$, left and $E=2$, right). The position of the interfaces is indicated
by vertical lines. The origin is taken at the $F_2/S$ interface. Only a
small part of the $S$ layer is shown, as the behavior of ${\bf S}$
is constant in $S$ beyond the region included.
In each panel, we  plot the results for seven values
of the angle $\phi$, as indicated by the key in the upper right
panel of Fig.~\ref{Fig2d}.   
In each case we see that the
spin currents at
$\phi=0$ and $\phi=180^{\circ}$ are constant, as there are no spin torques when ${\bf h_1}$ and ${\bf h_2}$ are collinear. 
Furthermore, $S_x$ for $\phi=90^{\circ}$ is constant in $F_2$ since
 ${\bf h_2}$ in this case is along the $x$-axis. 
Similarly $S_z$ is constant for all $\phi$ in $F_1$ since, with our
choice of coordinates, ${\bf h_1}$ is along the $z$-axis. 
As the bias increases, the magnitude of the spin current increases, 
except for the $y$-component, normal
to the layers, which is nearly bias independent. 
This is because  $S_y$ is driven primarily by the static 
spin torque that exists near the boundary
of the ferromagnetic layers: this
torque is entirely in the $y$-direction. 
We see that $S_y=0$ for all $\phi$ and all biases within the $S$ layer. 
This is possibly because the excess current in $S$ is
 due to triplets, and there are none formed in the $y$ direction.
On the other hand, the $S_x$ and $S_z$ components within 
the superconductor become nonzero  at high bias 
for all angles $\phi$. These nonzero spin currents,
in $S$, occur when the applied bias is greater than the critical bias
(CB). This bias corresponds
to a  value 
smaller than $\Delta_0$: it represents the 
effective gap energy that the superconductor provides near
the interface. It 
has a nonmonotonic dependence on $\phi$. 
This dependence of the CB is due to the proximity 
effect between the $F_2$ and $S$ layers. The angular dependence comes from the formation of triplet pairs where there is angular mismatch in the system.
In this  case, with perfect interfaces, the angular dependence of
 the CB is large, confirming
 previous results for 
the charge current\cite{Moen2017}. 
It can be observed
 that at $E=0.6$, the critical bias 
values for each angle are sometimes
above and sometimes below that value
of $E$. For angles such that the CB is greater than the 
bias ($E=0.6$ in this case), the
spin current is zero in the superconductor. However, when the CB is lower than the applied bias, 
the excitations have energy greater than the 
effective gap energy and at those angles we find
 non-zero spin current in $S$. 

By viewing the spin current in 3D, we can get a better grasp of its
overall  orientation within the multilayer. In Fig.~\ref{figure3D},
in the high bias limit and at $\phi=90^\circ$, we see that the spin 
current  rotates 
in the $x-z$ plane from near the $z$ direction in $F_1$ to an angle close to 
the mismatch angle $\phi$ in $F_2$ and $S$. 
In the ferromagnetic layers, we see the spin current precessing about the exchange fields ${\bf h_1}$ and ${\bf h_2}$ in $F_1$ and $F_2$ respectively. 
The precession in $F_2$, however, is damped due to the proximity effect of the superconductor, 
the current becoming constant at the $F_2/S$
boundary. The spin current in the normal metal layer is also constant, since there are no torques there. 
The orientation of the spin current in $N$ is rotated in the $x-z$ plane to an angle between $0$ and $\phi$, with a nonzero $y$-component 
that is due to the net STT in both ferromagnetic layers. 

In Fig.~\ref{Fig2b} we examine  the $x$ and $z$ components
of  $\delta{\bf m}$ for low 
to high biases (left to right) as functions of $Y$. 
The $y$-component  is several orders of magnitude smaller and we do not
show it.  The component 
$\delta m_x$  
is zero for  $\phi=0$ 
and $\phi=180^{\circ}$. $\delta m_z$ is nonzero 
and only weakly $\phi$ dependent
in $F_1$, whereas $\delta m_x$ is oscillatory and small 
in this region. Furthermore, $\delta m_z$ and $\delta m_x$ are 
nonzero and nearly constant with
position in the $S$ region at large bias. In general
 the magnitude of the spin accumulation is oscillatory 
everywhere at low biases, but with small amplitudes. 
The spin accumulation oscillates in $N$ and 
irregularly rotates in the $x-z$ plane, particularly for mismatch angles near 
 $\phi=90^{\circ}$. The overall magnitude increases with bias 
with very little change in the  angular dependence. The spin accumulation 
vector tends to align with ${\bf h_2}$ within the superconductor: this
is similar to the spin current behavior. The magnitude 
of $\delta \mathbf{m}$ also decreases, in all layers, as $\phi$ 
increases from $0$ to $180^{\circ}$.

In Fig.~\ref{Fig2c} we examine the 
spatial average (as defined earlier in
this section) of the spin accumulation in 
the $N$ and $S$  layers (upper and lower plots, respectively), as a function
of bias.  In both regions,  $\langle\delta m_x\rangle$ 
vanishes for $\phi=0$ and $\phi=180^{\circ}$. In $S$ we can see 
a critical bias behavior in $\langle\delta m_x\rangle$, at which 
value the magnitude begins to rise quickly with bias, becoming
approximately linear. In both regions each component is 
nonmonotonic in $\phi$. In $S$  $\langle\delta m_x\rangle$ is maximized between 
 $\phi=60^{\circ}$ and $\phi=90^{\circ}$ while in $N$ it is most
negative at $\phi=150^\circ$,
$\langle\delta m_z\rangle$ features a similar, but less dramatic critical bias feature
only in $S$, with this component decreasing for angles $\phi > 90^{\circ}$.

In Fig.~\ref{Fig2d} we consider the average spin transfer torques
as a function of $E$, as  just
done with the  average spin accumulation. We do so only in the 
ferromagnetic regions where the torques are nonzero. The component 
$\tau_z$ is zero in the outer ferromagnetic region $F_1$, since the field 
${\bf h_1}$ is along the $z$ direction, and it is not
plotted: the angular key for the entire figure is
shown instead. The torque ${\bf \tau}$ is always zero for  $\phi=0$ and $\phi=180^{\circ}$, and $\tau_x=0$ for $\phi=90^{\circ}$ in $F_2$: 
this follows from our geometry. We see 
a strong critical bias feature in the $x$ component in
both $F_1$ and $F_2$, and also in the $z$ components in 
$F_2$: the averaged torque is zero below the CB,
and then grows linearly with increasing bias. 
The $x$ component in $F_1$, and the $z$ component in $F_2$, show
 similar behavior,  with a steady increase or decrease in value
 respectively for all angles, and a maximum magnitude between $\phi=90^{\circ}$ and $\phi=120^{\circ}$. 
$\langle\tau_x\rangle$ in $F_2$ is different: it 
 increases with $E$ for angles $\phi < 90^{\circ}$ and decreases
 for angles $\phi > 90^{\circ}$. $\langle\tau_y\rangle$ has very different behavior from
both of the other components: it is nonzero at zero bias due to the static ferromagnetic proximity effect. Because of this, $\langle\tau_y\rangle$ is nearly independent 
of bias, slightly decreasing in magnitude in both ferromagnetic regions. 
It follows from Eqn.~\ref{spinconserve} in the steady state
that the net change in spin current in $N$ and $S$ is directly proportional to the average torque. Indeed, the constant $S_y$ in the normal metal can be described by the net average torque $\tau_y$ in both ferromagnetic regions.
Much of the above discussion for Fig.~\ref{figure2} will apply to the 
results for other physical parameter values
presented below. 

\begin{figure*}  
  \centering
  \vspace{-1cm}
  \begin{subfigure}[b]{.48\textwidth}
    \hspace*{-1.7cm}\includegraphics[angle=-90,width=1.3\textwidth] {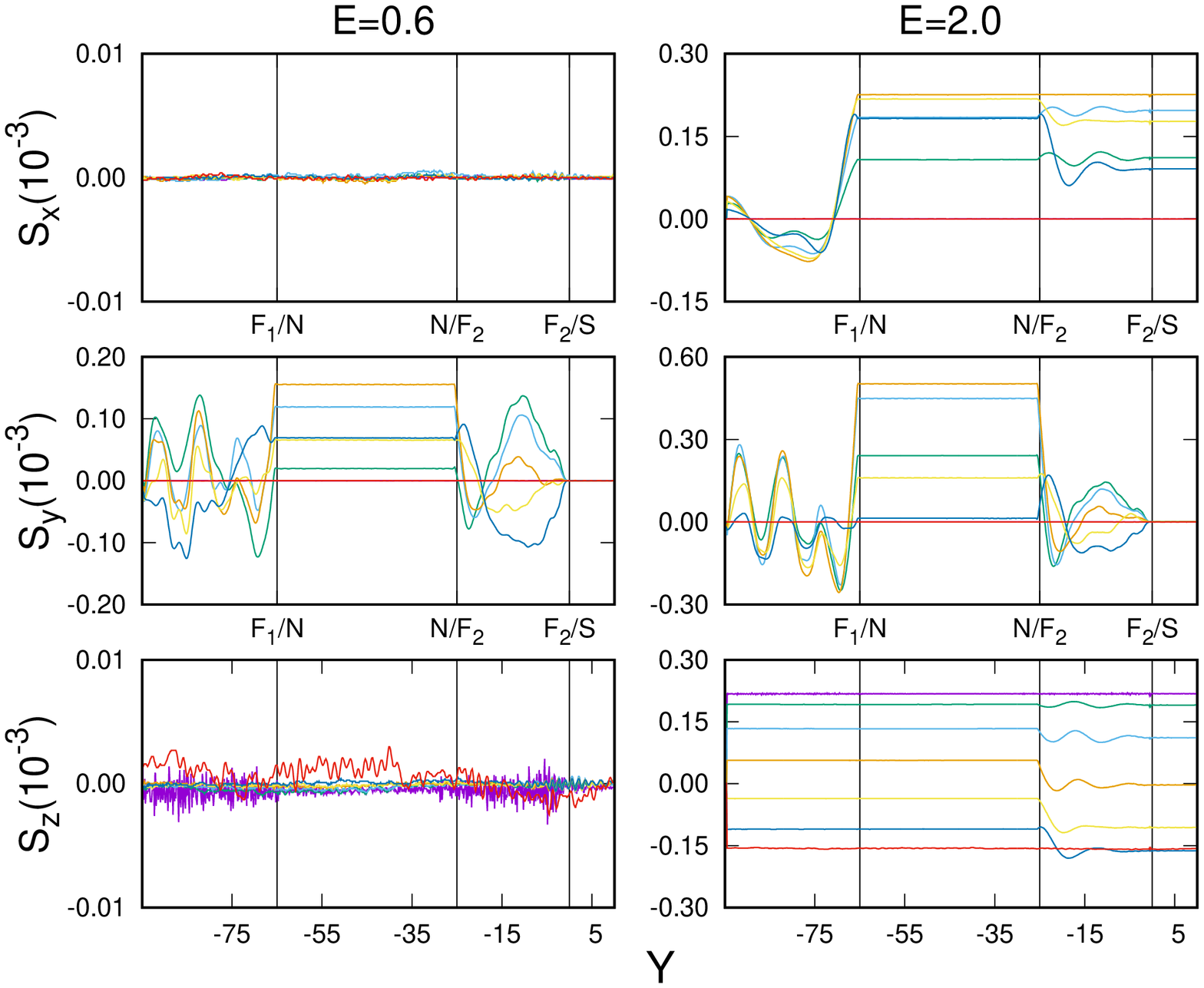} 
    \vspace*{-0.1cm}\small\caption{Local Spin Current}
    \label{Fig3a}
  \end{subfigure}
  \hfill
  \begin{subfigure}[b]{.48\textwidth}
    \hspace*{-1.1cm}\includegraphics[angle=-90,width=1.3\textwidth] {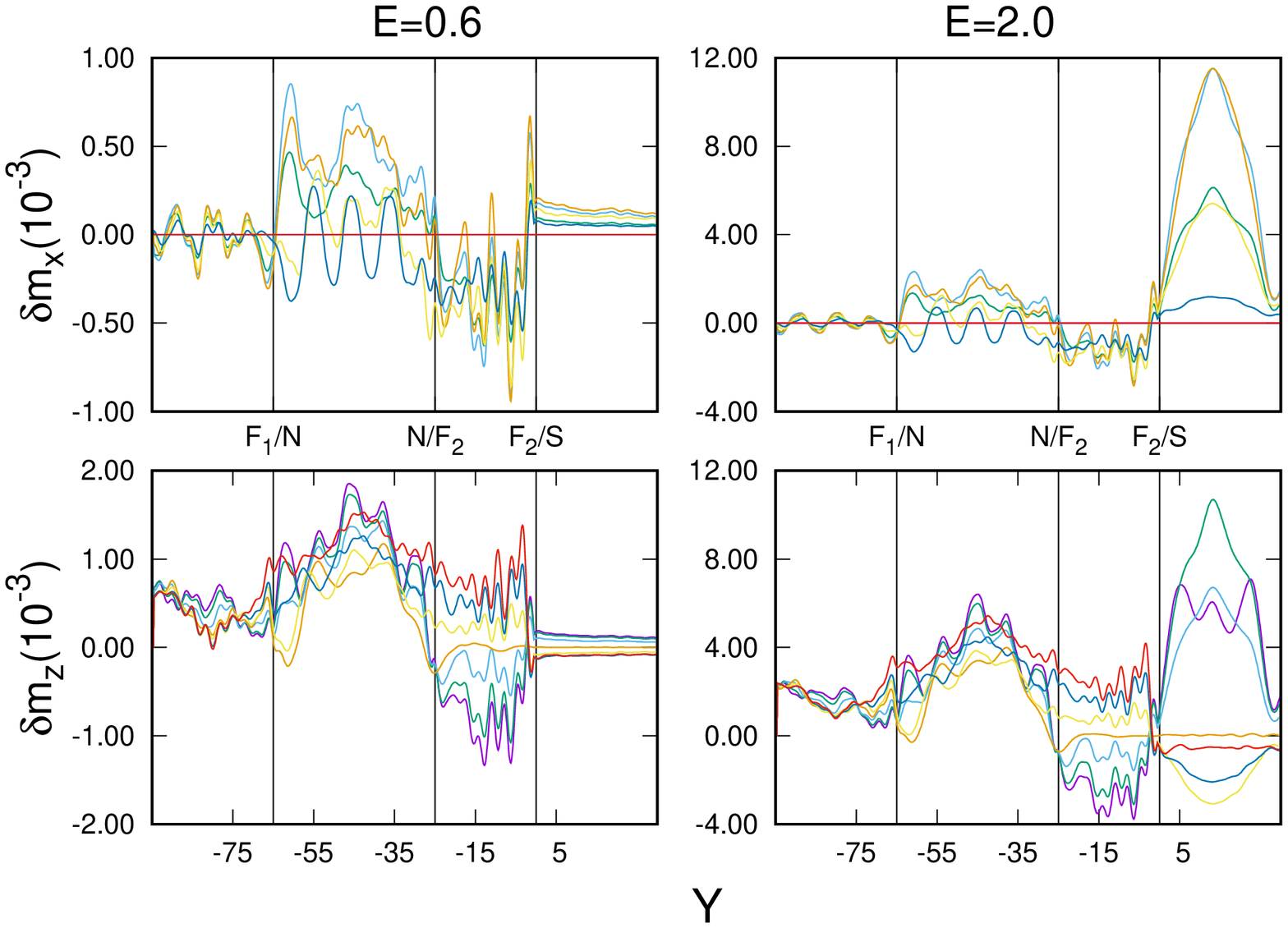}\vspace*{-0.6cm}
    \small\caption{Local Spin Accumulation}
    \label{Fig3b}
  \end{subfigure}
  \vskip\baselineskip
  \vspace*{-1.7cm}
  \begin{subfigure}[b]{.48\textwidth}
    \hspace*{-1.3cm}\includegraphics[angle=-90,width=1.3\textwidth] {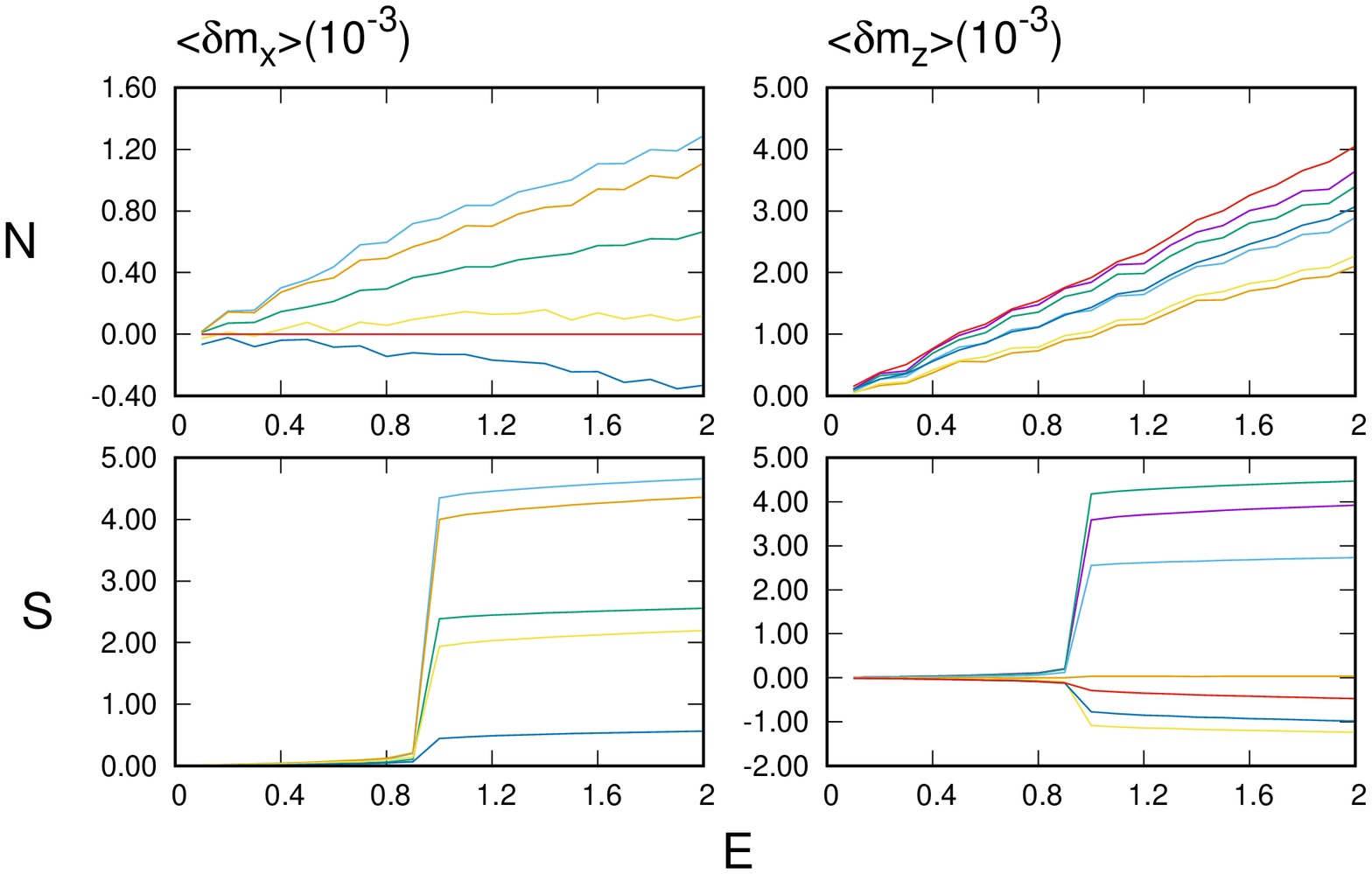} 
    \vspace*{-1.0cm}\small\caption{Spatially Averaged Local Spin Accumulation}
    \label{Fig3c}
  \end{subfigure}
  \quad
  \begin{subfigure}[b]{.48\textwidth}
    \hspace*{-1.7cm}\includegraphics[angle=-90,width=1.3\textwidth] {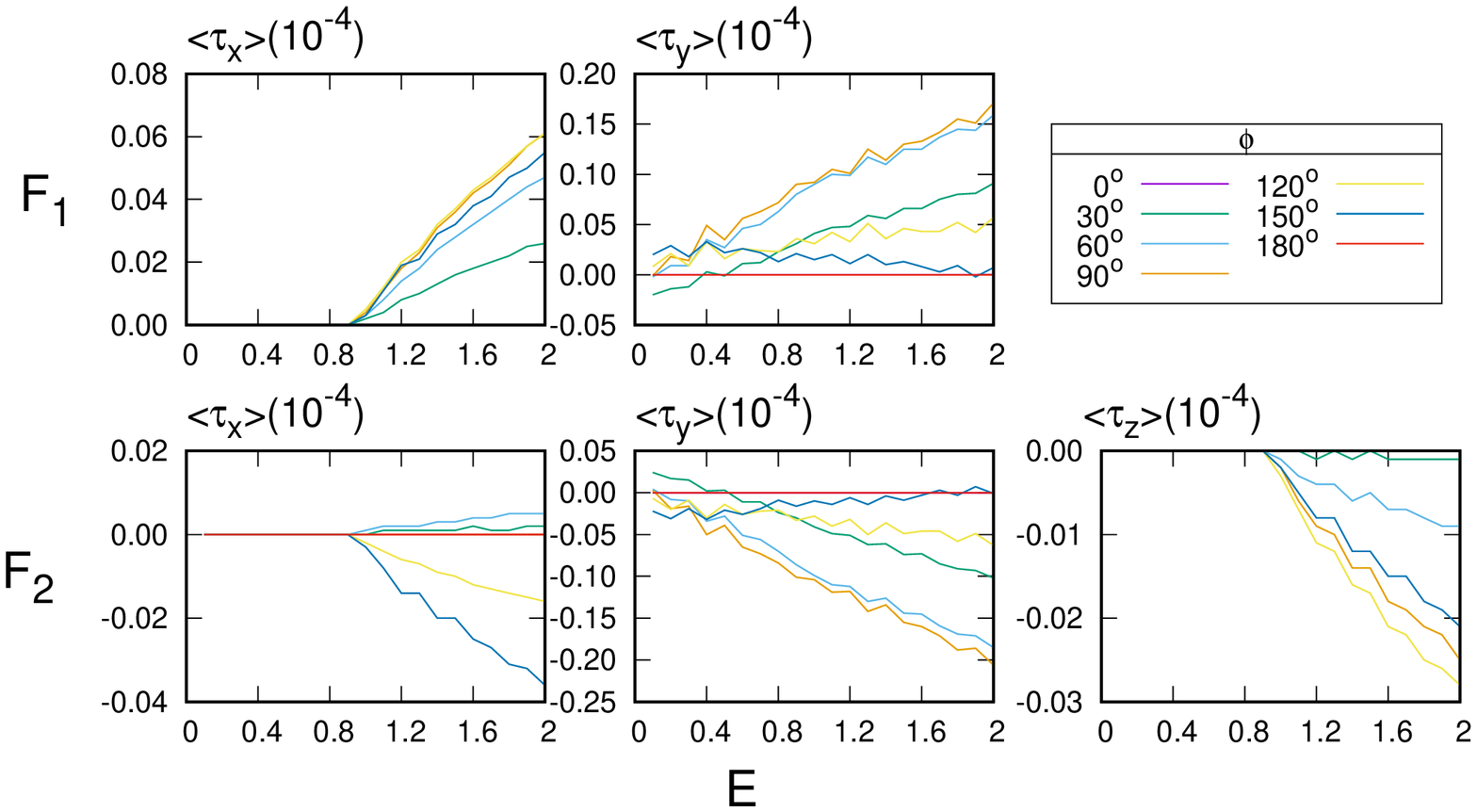}\vspace*{-1.7cm}
    \small\caption{Spatially Averaged Spin Torque}
    \label{Fig3d}
  \end{subfigure}
\caption{Results for a nonzero tunneling barrier at 
the $F_2/S$ interface. The layer thicknesses are as in Fig.~\ref{figure2}
 and the interfacial barriers are $H_B=0$ and $H_{B3}=0.9$. See text for details.}
\label{figure3}
\end{figure*}

\subsection{Interfacial Scattering}
We now turn on the effect of interfacial scattering. First we consider, 
in Fig.~\ref{figure3},  the case where only a barrier
at the $F_2/S$ interface exists,  with a qualitatively 
large scattering parameter value
$H_{B3}=0.9$. The layer thicknesses are as
in the previous figure. When the scattering is 
large at this interface, the superconducting
proximity effect is reduced. We compare this case to the zero scattering 
limit of Fig.~\ref{figure2} in order to examine closely how
the basic features of the proximity effect influence the
spin currents.
The organization of the panels in Fig.~\ref{figure3} is the same
as in Fig.~\ref{figure2}. 

In Fig.~\ref{Fig3a} we see that the $x$ and $z$ components of the spin current are 
now driven to zero, within numerical
precision, at low bias. This is due to
the increase in the CB  due to the barrier, which
weakens the proximity effect and thereby 
makes it more difficult for the Cooper pairs to propagate 
out of the superconductor  and convert to long ranged
triplets. The $y$ component, however, 
is still nonzero due to the static spin torques from the ferromagnetic proximity effect. 
Unlike in the other cases discussed, $S_y$ now
 increases significantly at higher biases, although not as dramatically as the other two components. In the high bias regime, 
the system returns to precessing about ${\bf h}$ in the ferromagnetic 
regions. $\mathbf{S}$ is also rotated about the $x-z$ plane, 
this time closer to the second ferromagnetic field ${\bf h_2}$ which 
is oriented at an angle $\phi$. The overall magnitude of the 
spin current is of course reduced by the  barrier.

\begin{figure*} 
  \centering
  \vspace{-1cm}
  \begin{subfigure}[b]{.48\textwidth}
    \hspace*{-1.7cm}\includegraphics[angle=-90,width=1.3\textwidth] {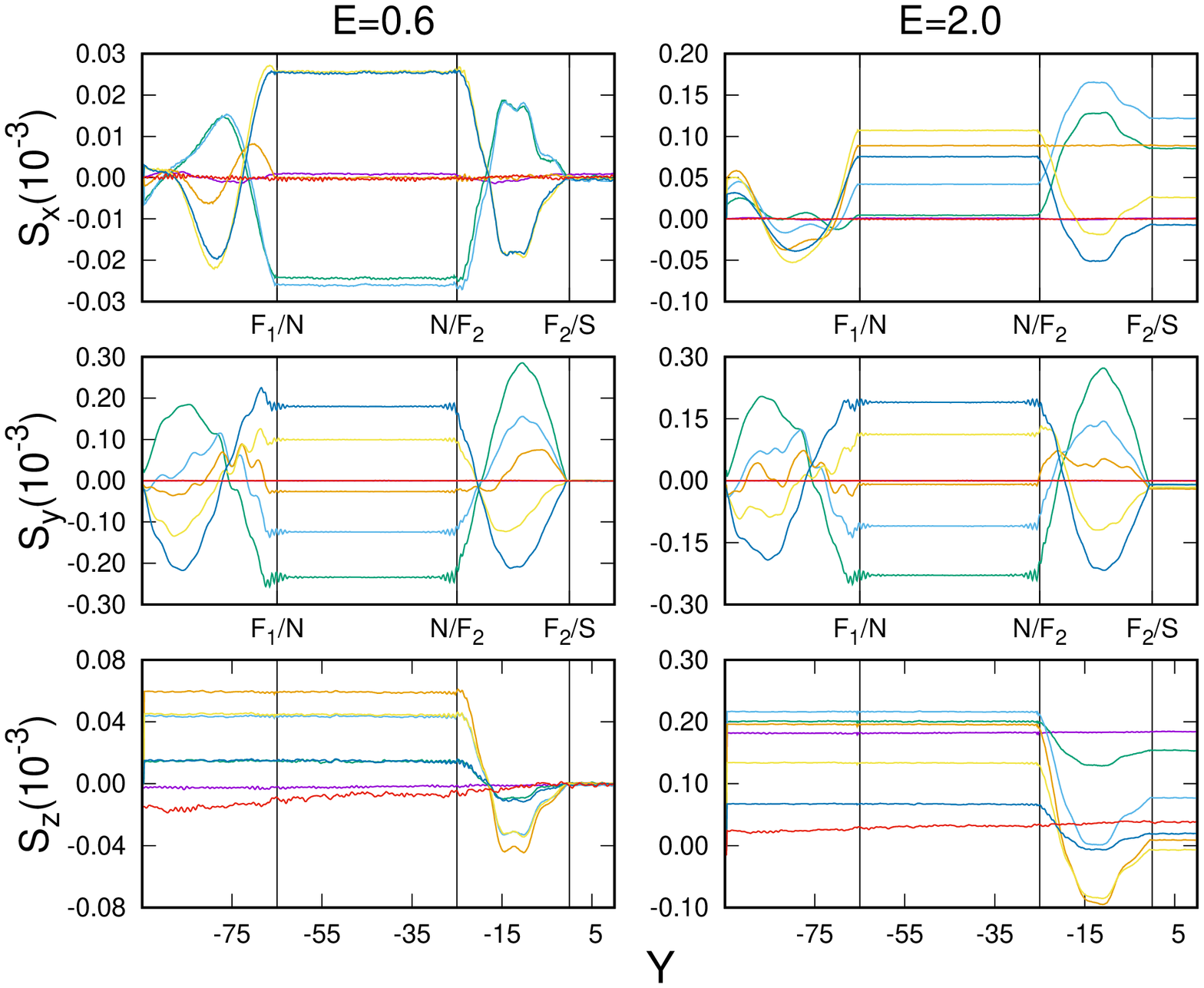} 
    \vspace*{-0.1cm}\small\caption{Local Spin Current}
    \label{Fig4a}
  \end{subfigure}
  \hfill
  \begin{subfigure}[b]{.48\textwidth}
    \hspace*{-1.1cm}\includegraphics[angle=-90,width=1.3\textwidth] {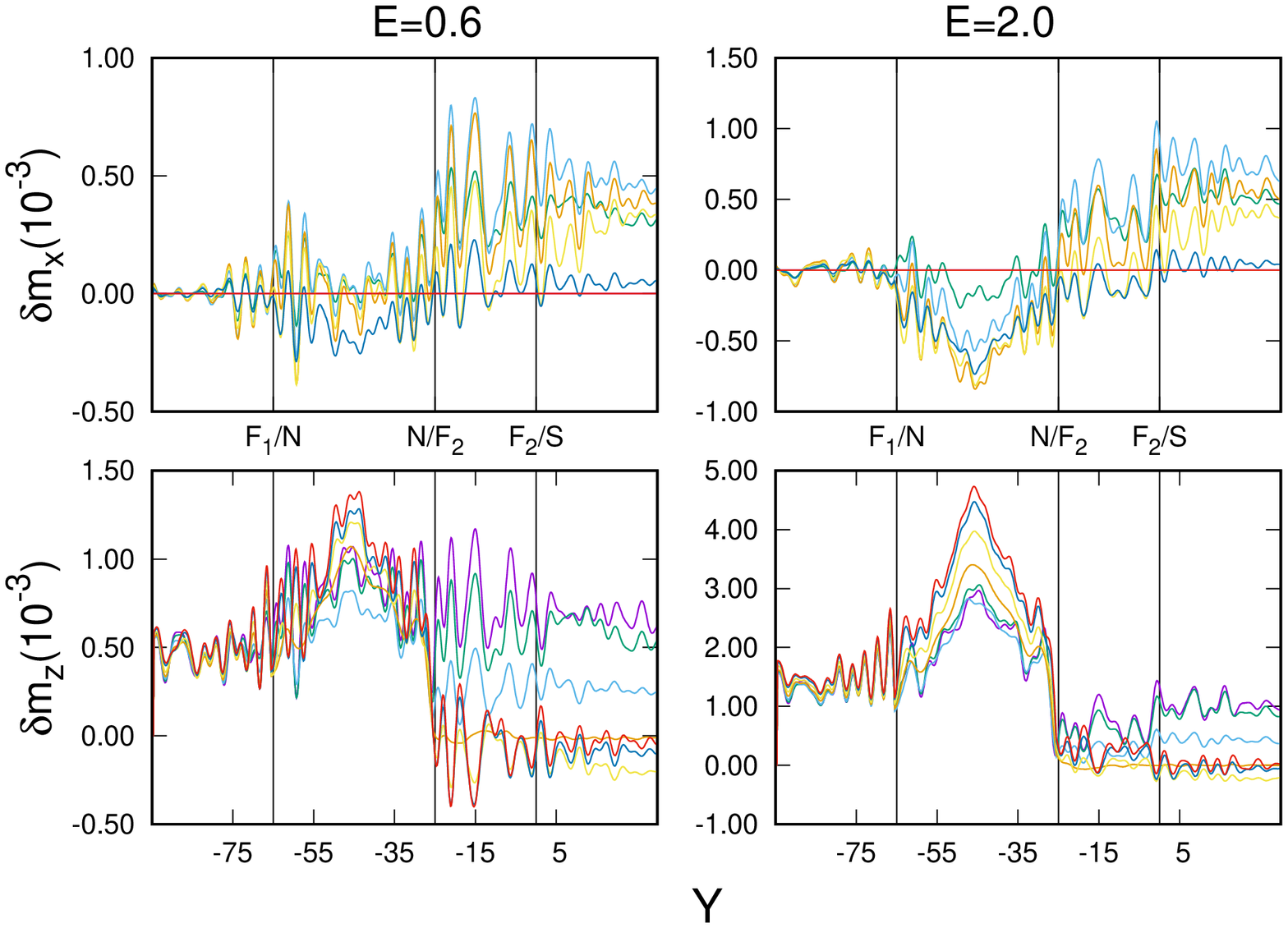}\vspace*{-0.6cm}
    \small\caption{Local Spin Accumulation}
    \label{Fig4b}
  \end{subfigure}
  \vskip\baselineskip
  \vspace*{-1.7cm}
  \begin{subfigure}[b]{.48\textwidth}
    \hspace*{-1.3cm}\includegraphics[angle=-90,width=1.3\textwidth] {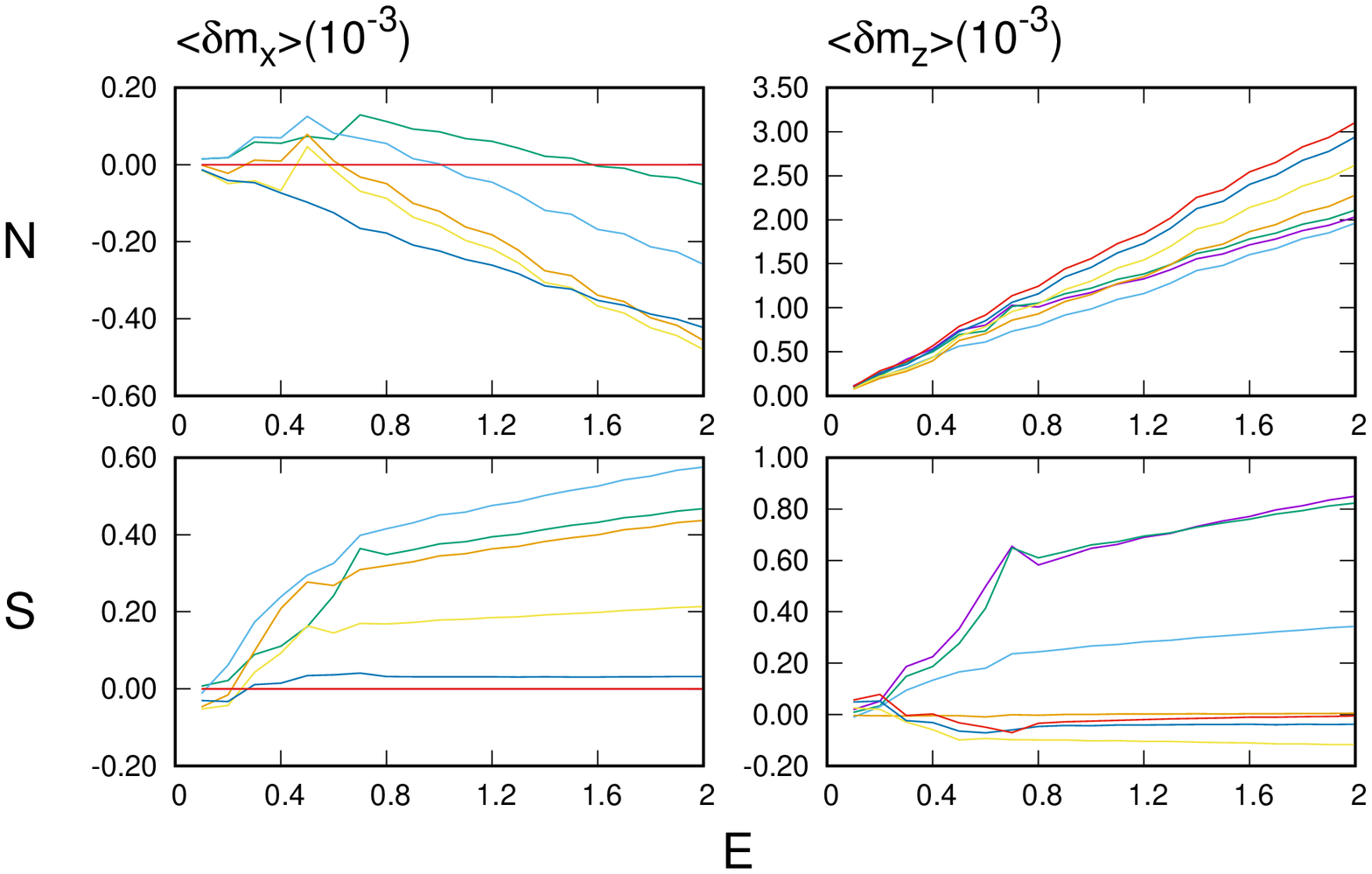} 
    \vspace*{-1.0cm}\small\caption{Spatially Averaged Local Spin Accumulation}
    \label{Fig4c}
  \end{subfigure}
  \quad
  \begin{subfigure}[b]{.48\textwidth}
    \hspace*{-1.7cm}\includegraphics[angle=-90,width=1.3\textwidth] {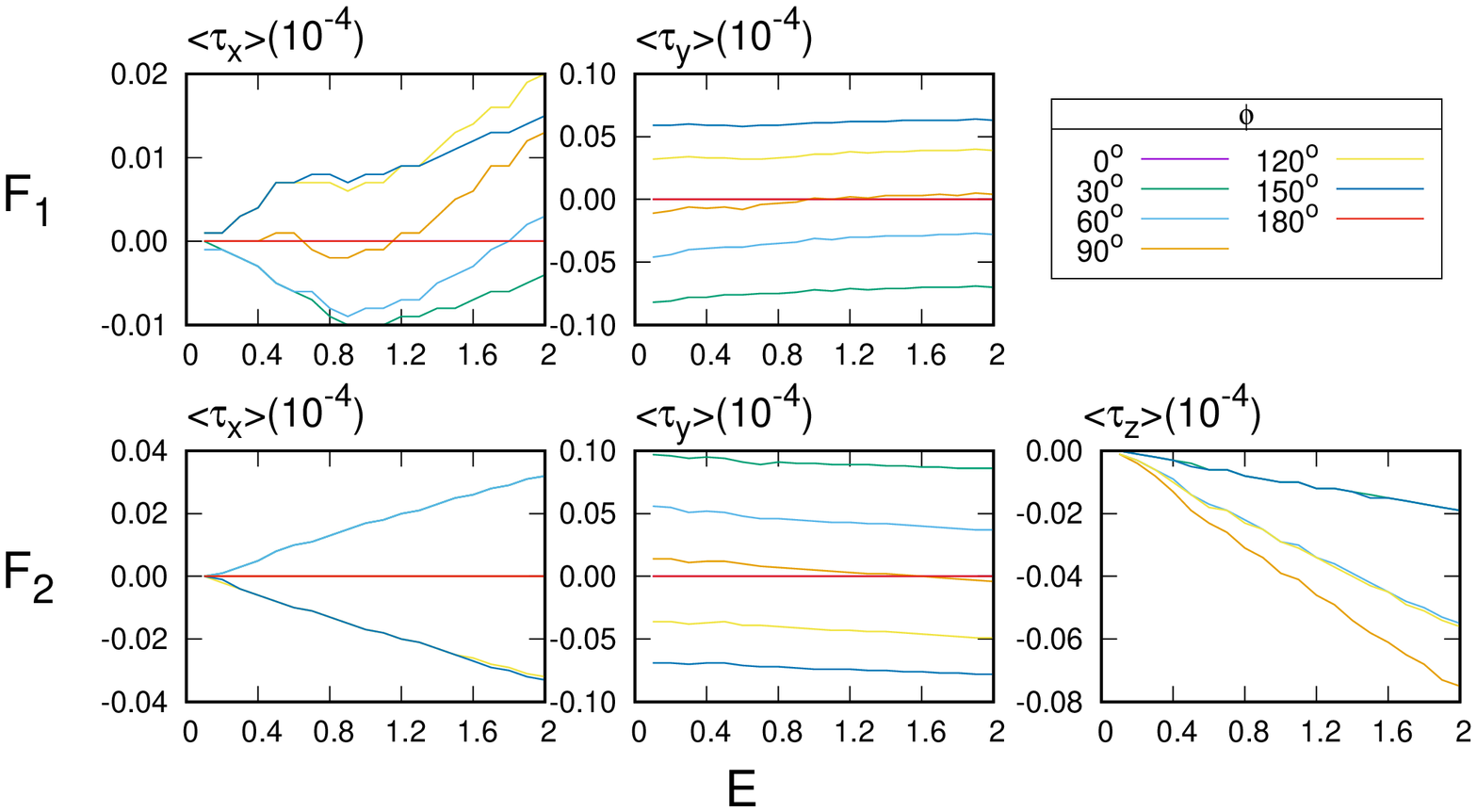}\vspace*{-1.7cm}
    \small\caption{Spatially Averaged Spin Torque}
    \label{Fig4d}
  \end{subfigure}
\caption{Results for nonzero barriers in the  $F_1/N$
and $N/F_2$ interfaces. The layer thicknesses  are
as in Figs.~\ref{figure2} and \ref{figure3}. The interfacial barriers 
are $H_B=0.5$ and $H_{B3}=0$. See text for details.}
\label{figure4}
\end{figure*}

In Fig.~\ref{Fig3b}  we see that the spin accumulation is 
significantly decreased in magnitude within the superconductor 
at the low bias limit. The magnitude increases dramatically 
in $S$ at  high bias, although it remains smaller
than for perfect interfaces.  Furthermore, we see that the 
magnitude of $\delta \mathbf{m}$ is highly oscillatory in the superconductor. The orientation remains fixed to that of the 
exchange field ${\bf h_2}$. In the normal metal, 
the spin accumulation rotates 
counterclockwise within the $x-z$ plane  for $\phi < 90^{\circ}$ and 
then reverses direction to become aligned with the $z$ axis 
again for $\phi=180^{\circ}$. The rotation in the $x-z$ plane is uniform throughout the $N$ layer in the high bias case, but not for low bias values.
In the spatially averaged results
of Fig.~\ref{Fig3c} we note a remarkable feature in the 
superconducting layer:
a  dramatic, sharp increase in the magnitude 
of $\langle\delta \mathbf{m}\rangle$ at the critical bias, 
after which the magnitude grows at a much slower rate. 
The angular dependence remains approximately the same as
in Fig.~\ref{Fig2c}. The low bias spin accumulation is heavily impeded by the high barrier.
In Fig.~\ref{Fig3d}  we show that the average STT exhibits
the  same critical bias features as in Fig.~\ref{Fig2d}. 
However, the high barrier causes the critical bias to increase
and to become nearly $\phi$ independent.  Its value is seen to be
 $E\approx 0.85$ in the results for $\langle\tau_x\rangle$ 
(in both $F_1$ and $F_2$)
and for $\langle\tau_z\rangle$ in $F_2$. Furthermore, $\langle\tau_x\rangle$ 
in $F_2$ shifts to 
become almost entirely negative. The $y$ component is changed 
dramatically by the barrier: $\langle\tau_y\rangle$  steadily increases
 in magnitude with increased bias for all angles except 
$\phi=150^{\circ}$. The static spin torque is heavily reduced by the introduction of a large barrier between $F$ and $S$, 
which increases the pair potential at  the interface.

In Fig.~\ref{figure4} we turn to the converse case where the
scattering potentials at both of the  $F/N$ interfaces
are nonzero, while the
$F_{2}/S$ barrier is ideal, thereby complementing the study in the previous
figure. The layer thicknesses  are again
$30/40/25/180$. For the interfacial barriers we take $H_B=0.5$
(a value not so high as
to be in the tunneling limit) and $H_{B3}=0$. Thus, there
is a full proximity effect between $S$
and $F_2$.  We now are interested in how the scattering within
the spin valve structure affects 
 the spin transport. Perhaps unsurprisingly, the introduction of these barriers turns out to  be very important, 
as the spin-valve effect, which determines much of the spin-transport features, is quite sensitive to these scattering potentials. 
In Fig.~\ref{Fig4a}  we see that the spin current is nonzero in the $N$ region at low bias, as in the zero barrier case. 
$S_y$ in $N$ is now almost entirely bias independent and 
its angular dependence is symmetric about $\phi=90^\circ$,
positive 
for  $\phi > 90^{\circ}$  and negative
for $\phi < 90^{\circ}$. Similarly, the $\phi$ 
dependence of $S_x$ at low bias is  nearly 
symmetrical with respect to $\phi$ in all layers. 
At high bias, we again see that the $x$ and $z$ components
of the spin current increase, penetrating the superconductor. 
Due to the significant interfacial scattering, the overall magnitude decreases from the zero barrier case, especially for the $x$ and $z$ components.
 
\begin{figure} 
\vspace*{-1.4cm}\hspace*{-1.cm}\includegraphics[angle=-90,width=0.63\textwidth] {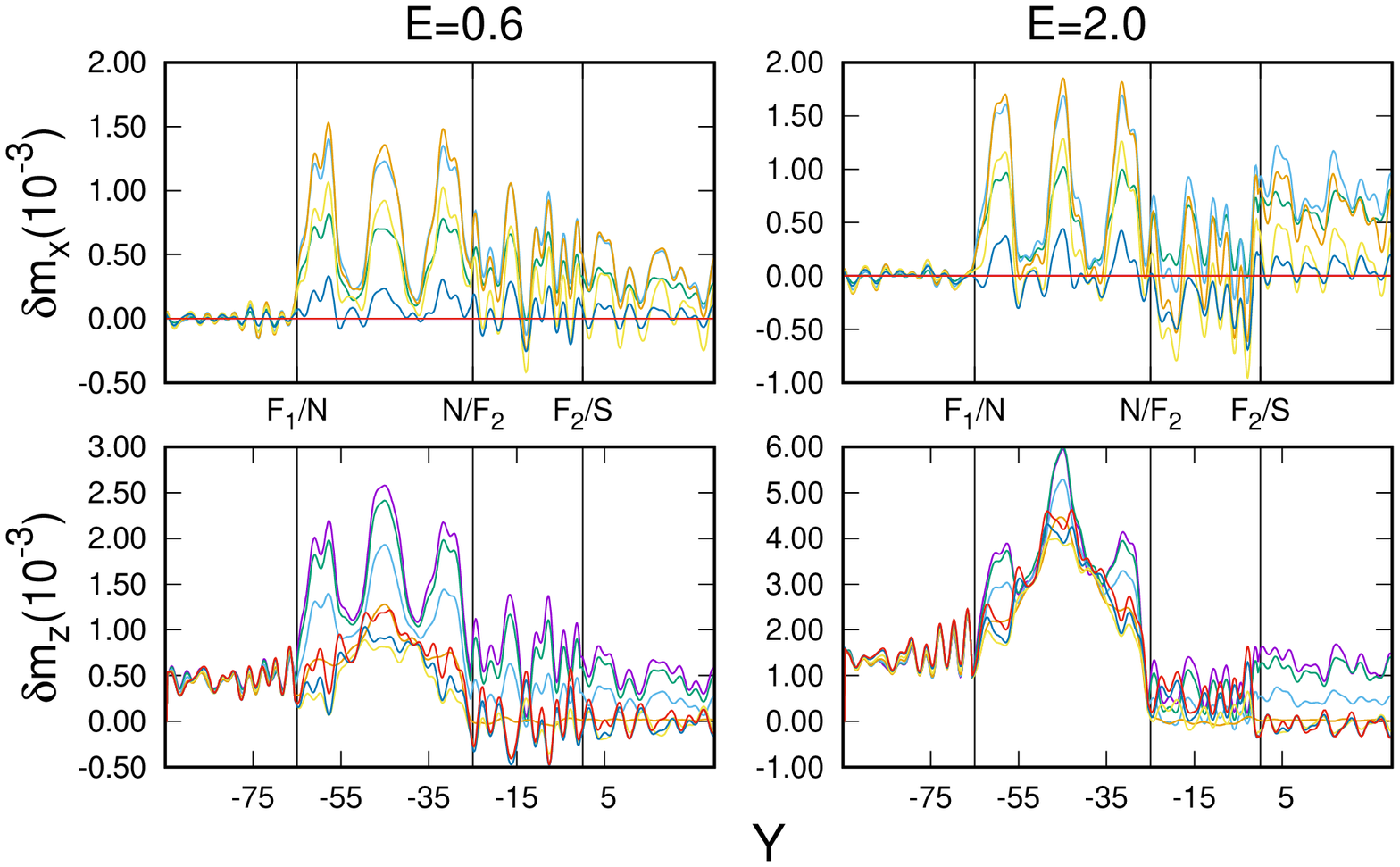}
\vspace*{-2.6cm}

\hspace*{-1.cm}\includegraphics[angle=-90,width=0.63\textwidth] {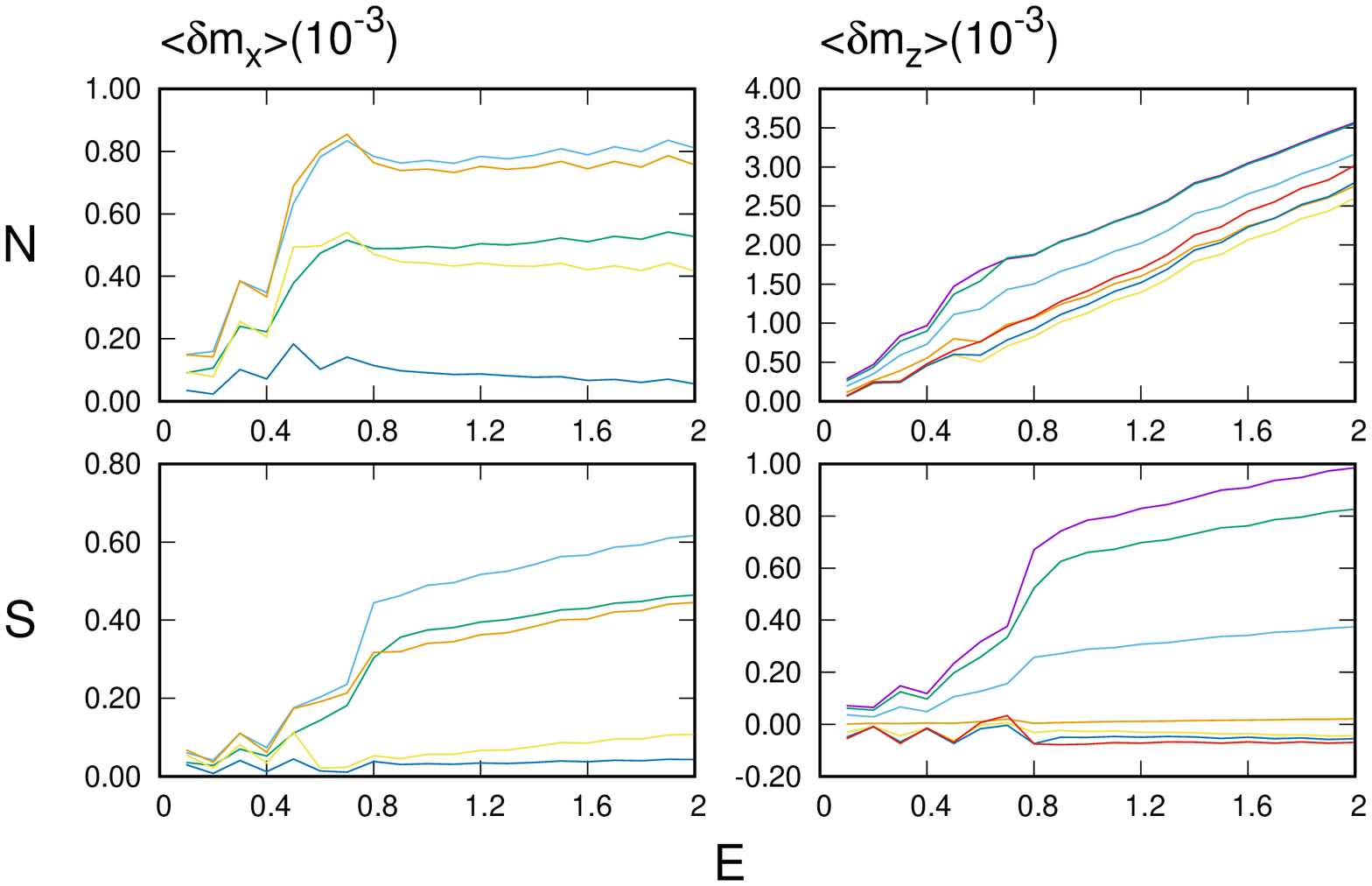}
\vspace*{-0.9cm}

\caption{ Results with  nonzero 
interfacial barriers at all interfaces. The layer thicknesses 
are as in the previous figures, and the interfacial 
barriers are $H_B=0.5$ and $H_{B3}=0.3$. The top four panels are the 
local spin accumulation, and the bottom four panels
are the spatially averaged spin accumulation. The color key for the
angular dependence is as in Fig.~\ref{Fig2d}. See text for details. 
}
\label{figure5}
\end{figure}

In Fig.~\ref{Fig4b} we see that, in comparison
to the corresponding
perfect interface  case of Fig.~\ref{Fig2b}, 
the angular dependence is decreased in the normal metal layer, 
with more oscillations in $\delta m_{x}$ about the zero 
value and a peak forming in $\delta m_z$ in both the low and high 
bias cases. 
In Fig.~\ref{Fig4c}  we see that the average
 spin accumulation in $S$ has an angular dependence and critical bias features similar to those
found in the zero 
barrier case, but with decreased magnitude. An exception is
 for the $x$ component at $\phi=150^{\circ}$, which is significantly 
larger. In the normal metal, $\langle\delta m_{x}\rangle$ increases up to 
a $\phi$ dependent CB, then steadily decreases for increasing bias. 
$\langle\delta m_z\rangle$ monotonically increases with bias, and has a greater magnitude than $\langle\delta m_x\rangle$.
In Fig.~\ref{Fig4d}  we see significant differences in the behavior of the average STT, as compared to the single high barrier case
of Fig.~\ref{Fig3d}. $\langle\tau_x\rangle$ in $F_1$ no longer features a CB behavior: it is nearly constant
with $E$. In both ferromagnets, $\langle\tau_y\rangle$ is again only
weakly dependent on bias, with a slight increase in the $F_1$ layer 
and a decrease in the $F_2$ layer. 
The overall magnitude is significantly smaller, in all layers and for
 all components, than in the zero barrier case. In $F_2$, we see
 a remarkable symmetry emerge in the angular dependence 
of the averaged $\tau_x$ and $\tau_z$. For $\langle\tau_x\rangle$, the values
 for $\phi=30^{\circ}$ and $\phi=60^{\circ}$ are both
 increasing and positive, while  
those for $\phi=120^{\circ}$ and $\phi=150^{\circ}$ are 
decreasing by an equivalent amount. 
Similarly, for $\langle\tau_z\rangle$, we see an equivalent  
decrease in value with increasing bias for supplementary angles ($\phi=30^{\circ},150^{\circ}$ and $\phi=60^{\circ},120^{\circ}$).

In Fig.~\ref{figure5}  we finally examine the 
relevant situation  where there are scattering barriers at all
interfaces. Thus, in addition
to the  two interfacial scattering barriers with $H_B=0.5$ in Fig.~\ref{figure4} 
we include an additional scattering barrier at the $F_2/S$ interface,
with $H_{B3}=0.3$. Although
it is reasonable to assume  that efforts will  be made to
 minimize the scattering at this interface, unavoidable 
experimental limitations and wavevector mismatch 
(as mentioned above)  
imply that one can never assume that any barrier
will perfectly vanish. The layer thicknesses are 
as in the previous figures. The organization
of this figure is simplified, when compared to the
previous ones.  
The local spin current 
is not shown in Fig.~\ref{figure5} because it
 is very similar to that  in Fig.~\ref{Fig4a}. 
We see then that the  introduction of a 
third barrier of intermediate size 
at the $F_2/S$ interface does not significantly 
affect the spin current. The spin transfer torques also remain 
unaffected: this is because the proximity effect is not 
seriously inhibited by this additional barrier, 
and the spin-valve effect dominates the spin transport, in these cases.
Hence, the sets of panels corresponding to (a) and (d) in
the previous figures are omitted, and  
we focus in this figure on the spin accumulation 
and its spatial average, panels (b) and (c)
in the previous figures, now  in the top four and bottom four panels 
respectively. The color key for the $\phi$ dependence is as indicated
in Figs.~\ref{Fig2d} and \ref{Fig3d}.

In the top panels we see that $\delta \mathbf{m}$  in the normal metal 
layer departs significantly from what we found 
in Fig.~\ref{Fig4b} at $H_{B3}=0$. In $\delta m_z$ we 
observe a transition from the single peak result seen in 
Fig.~\ref{Fig4b} to a triple peak structure particularly
prominent for $\phi < 90^{\circ}$. The $x$ component also forms 
three peaks at low and high biases in $N$, at all angles. As
in the previous cases, $\delta \mathbf{m}$ is rotated in 
the $x-z$ plane in $N$. However, these rotations are non-uniform, and 
strongly non sinusoidal, with the troughs aligning with the $z$ axis 
while the peaks align at an angle less than the mismatch angle $\phi$.

\begin{figure} 

\vspace*{-1.4cm}\hspace*{-1.cm}\includegraphics[angle=-90,width=0.63\textwidth] {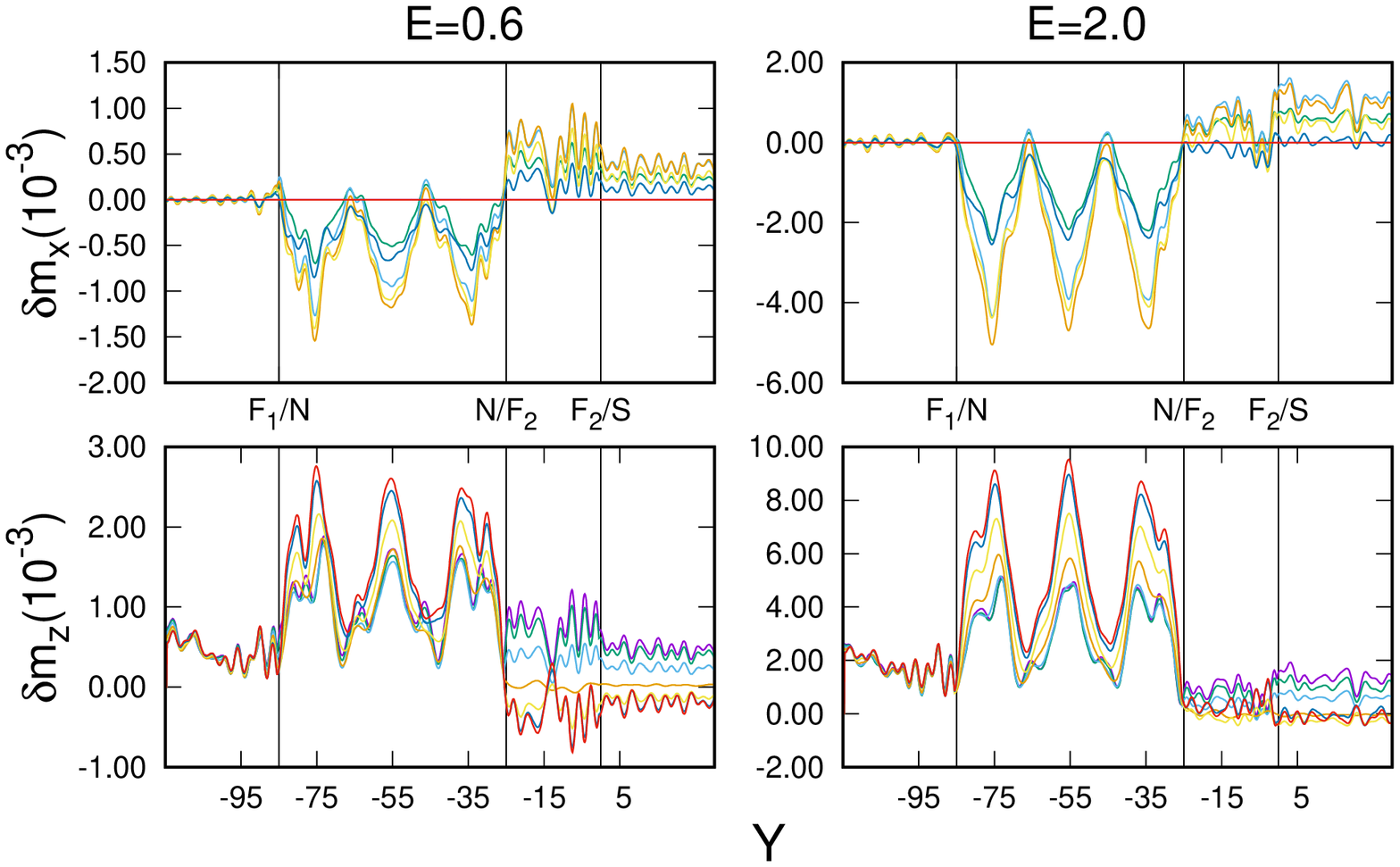}
\vspace*{-2.6cm}

\hspace*{-1.cm}\includegraphics[angle=-90,width=0.63\textwidth] {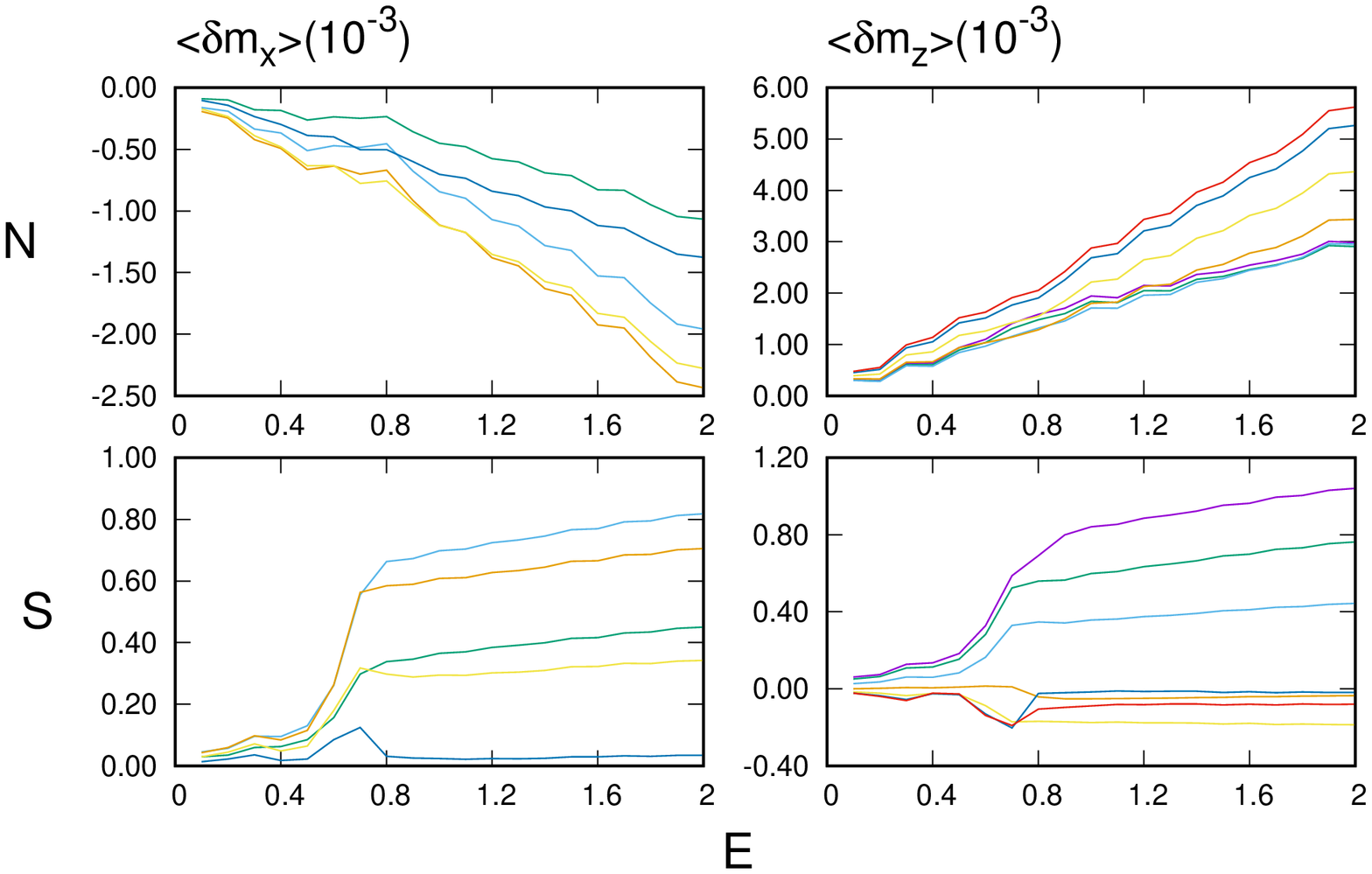}
\vspace*{-0.9cm}
 
\caption{ Results with an increased normal metal layer thickness, emphasizing the $D_N$ dependence. The layer thicknesses for the $F_1/N/F_2/S$ layers are
$30/60/25/180$ respectively, and the interfacial barriers $H_B$ and $H_{B3}$ are $0.5$ and $0.3$ respectively. The top four panels are the 
local spin accumulation, and the bottom four panels
are the spatially averaged spin accumulation. The color key for
the $\phi$ dependence is as in e.g. Fig.~\ref{Fig2d}.  
}
\label{figure6}
\end{figure}
In the bottom panels we see, in $\langle \tau_i \rangle$  an enhancement in the critical bias feature in $S$ 
seen in Fig.~\ref{Fig4c}, reflecting
that  the low bias conductance is depressed in this case\cite{Moen2017}. 
There is a steep growth in the magnitude of $\delta \mathbf{m}$,
averaged in $S$, at 
the critical bias. In the normal metal, we see a  behavior 
for $\langle\delta m_z\rangle$ similar to that in Fig.~\ref{Fig4c} but with a 
remarkably different  angular dependence. 
For $\langle\delta m_x\rangle$ in $N$ we see a very different high bias behavior, 
where $\langle\delta m_x\rangle$ increases dramatically at the critical bias and then abruptly 
levels off to a flat
or slightly decreasing bias dependence. The behavior in the average
$\delta m_x$ in $N$ is now much more similar to that of $\delta m_x$ or 
$\delta m_z$ in $S$.

\subsection{Dependence on Layer Thickness}

\begin{figure*} 
  \centering
  \vspace{-1cm}
  \begin{subfigure}[b]{.48\textwidth}
    \hspace*{-1.7cm}\includegraphics[angle=-90,width=1.3\textwidth] {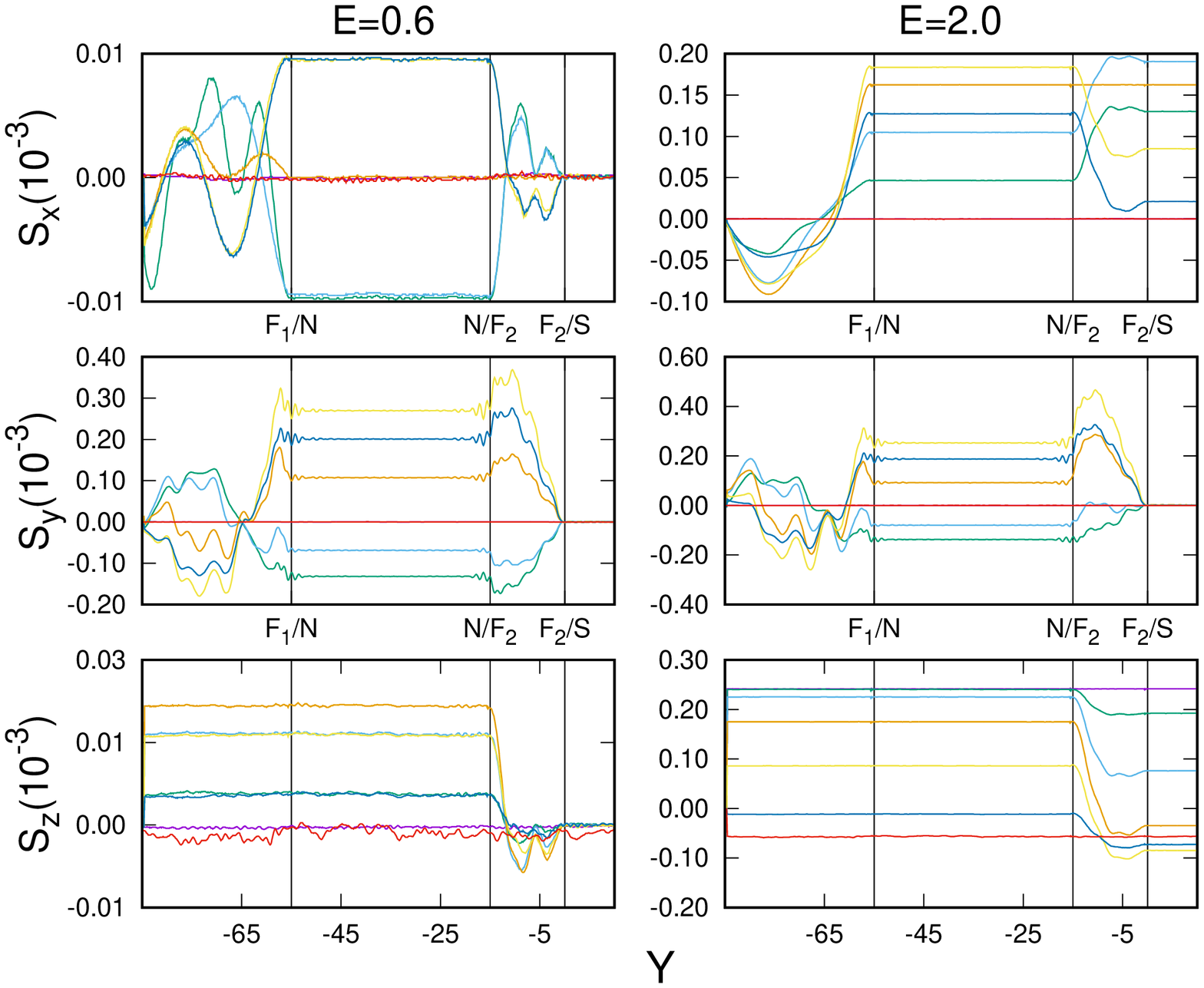} 
    \vspace*{-0.1cm}\small\caption{Local Spin Current}
    \label{Fig7a}
  \end{subfigure}
  \hfill
  \begin{subfigure}[b]{.48\textwidth}
    \hspace*{-1.1cm}\includegraphics[angle=-90,width=1.3\textwidth] {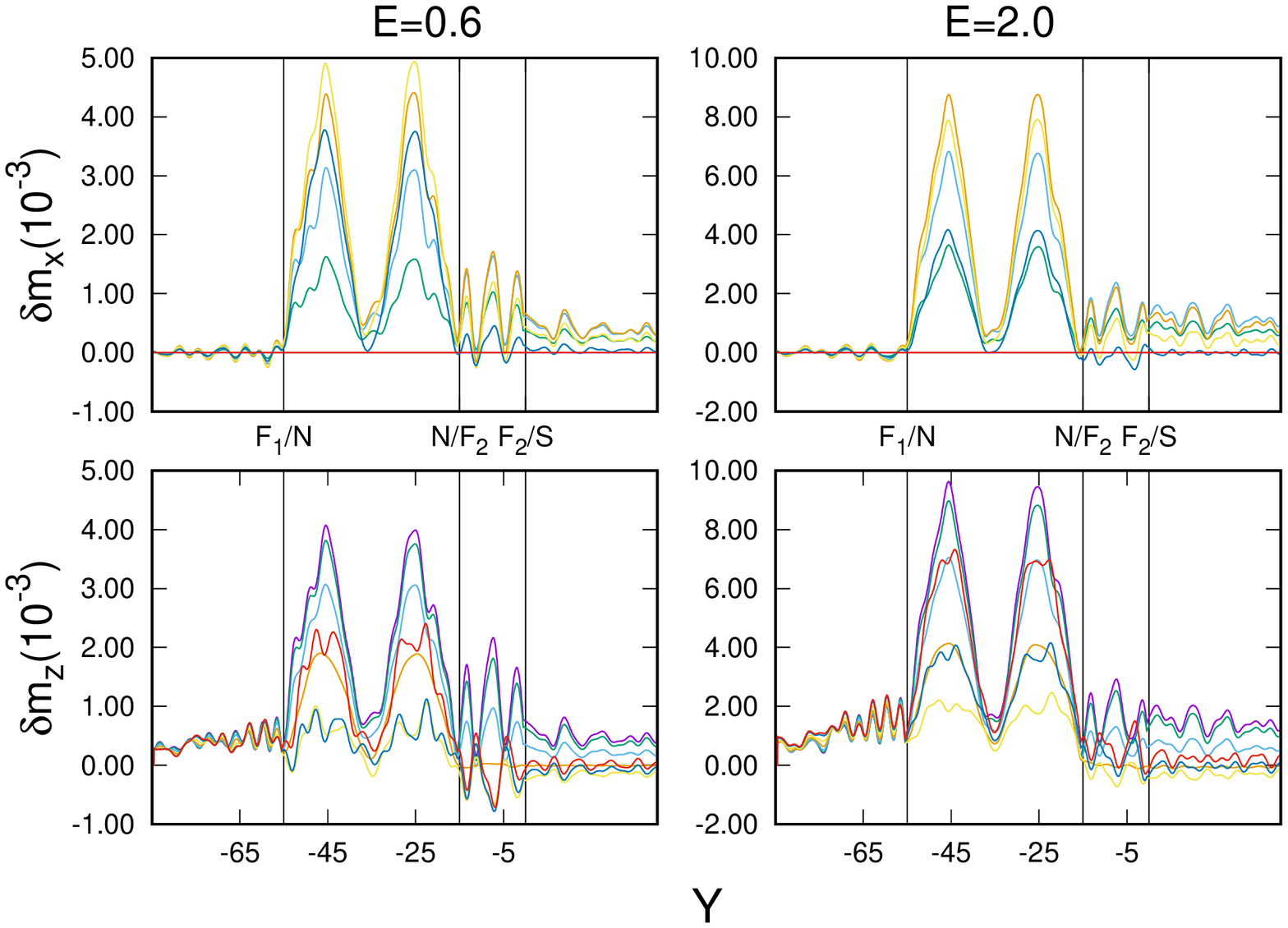}\vspace*{-0.6cm}
    \small\caption{Local Spin Accumulation}
    \label{Fig7b}
  \end{subfigure}
  \vskip\baselineskip
  \vspace*{-1.7cm}
  \begin{subfigure}[b]{.48\textwidth}
    \hspace*{-1.3cm}\includegraphics[angle=-90,width=1.3\textwidth] {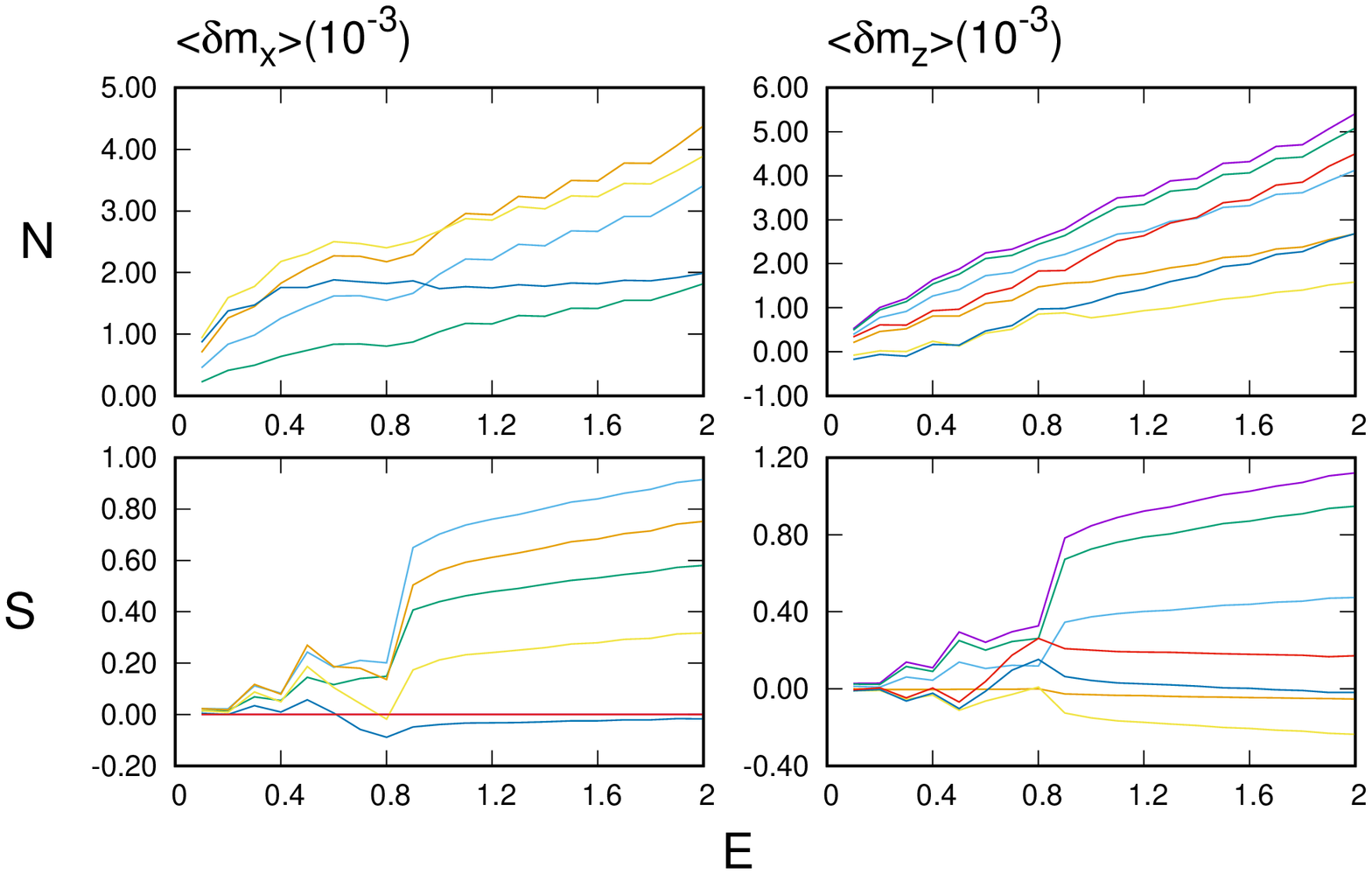} 
    \vspace*{-1.0cm}\small\caption{Spatially Averaged Local Spin Accumulation}
    \label{Fig7c}
  \end{subfigure}
  \quad
  \begin{subfigure}[b]{.48\textwidth}
    \hspace*{-1.7cm}\includegraphics[angle=-90,width=1.3\textwidth] {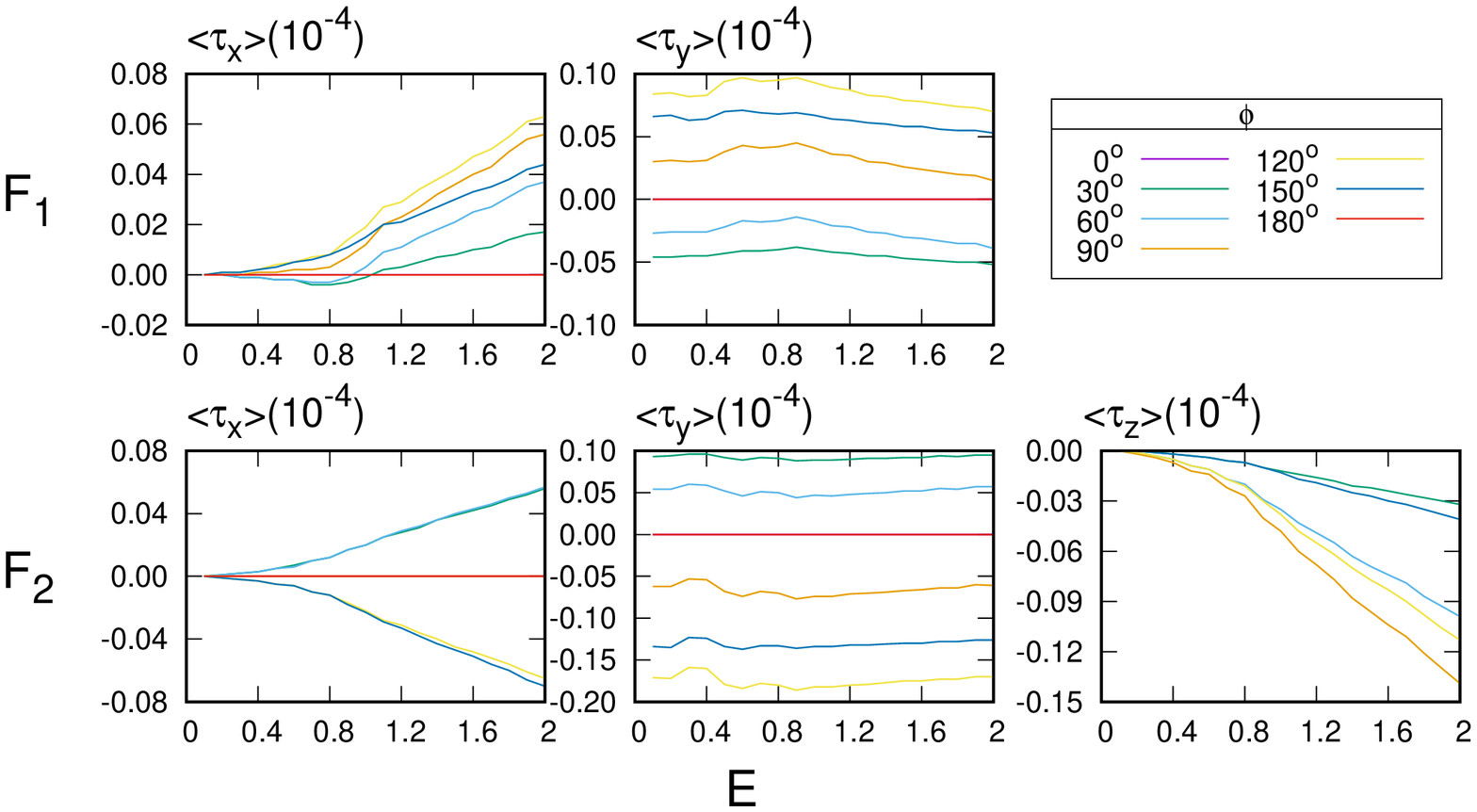}\vspace*{-1.7cm}
    \small\caption{Spatially Averaged Spin Torque}
    \label{Fig7d}
  \end{subfigure}
\caption{Results with a decreased intermediate ferromagnetic layer thickness, emphasizing the $D_{F2}$ dependence. The layer thicknesses for the $F_1/N/F_2/S$ layers are
$30/40/15/180$ respectively, and the interfacial barriers 
are $H_B=0.5$ and $H_{B3}=0.3$.}
\label{figure7}
\end{figure*}

In the next two figures,
Fig.~\ref{figure6} and Fig.~\ref{figure7}, we consider the  
dependence of the results  on geometry, i.e. on layer thickness. We
examine a situation where tThe top four panels are the 
local spin accumulation, and the bottom four panels
are the spatially averaged spin accumulationhe scattering
barriers are all nonzero and have
the same values as in Fig.~\ref{figure5}, namely $H_B=0.5$
and $H_{B3}=0.3$, but we now vary the 
intermediate layer thicknesses of the normal
metal, $D_N$, (Fig.~\ref{figure6}) and then  that of
the the inner ferromagnet,
$D_{F2}$ (Fig.~\ref{figure7}). The layer thicknesses of the $F_1$ and $S$  layers 
remain $D_{F1}=30$ and $D_S=180$ in both figures.  
In Fig.~\ref{figure6} we increase the normal metal layer spacing 
from the previous value $D_N=40$ to $D_N=60$, leaving $D_{F2}=25$, while in
 Fig.~\ref{figure7} we decrease the inner ferromagnetic layer 
thickness from $D_{F2}=25$ to 
$D_{F2}=15$, while leaving $D_N=40$. 
Geometric changes can strongly affect the 
transmission and reflection 
amplitudes, just as they do in  elementary quantum mechanics problems
such as that of transmission across two barriers, 
where the results can depend drastically on the separation
 between the two scattering centers. 
Here we examine how these rather minor changes in the geometry affect
 the spin-transport quantities. 
We have found little   change in the spin current and spin torque 
when increasing $D_N$, thus in Fig.~\ref{figure6} we only include 
plots of the spin accumulation and its average, following the scheme
of Fig.~\ref{figure5}, 
 in the top four and bottom four panels respectively. For Fig.~\ref{figure7},
on the other hand, we include the results for spin current and
torque components as we find nontrivial changes in the magnitude and orientation of the spin current, following then the organizational scheme of
Figs~\ref{figure2}, \ref{figure3}, and \ref{figure4}.

In the top panels of Fig.~\ref{figure6} we observe a three peak 
structure for the spin accumulation in $N$ similar to that found
 in the top panels of Fig.~\ref{figure5}, but with several
 distinctions. First, we see that $\delta m_z$ has now
fully transitioned to the three peak behavior for all $\phi$ and all 
biases. Also, the three peak behavior is  inverted in $\delta m_x$. 
Indeed, $\delta \mathbf{m}$ makes  now a clockwise rotation
in the $x-z$ plane in $N$, contrary to both the spin current and 
spin accumulation behaviors we have seen thus far. 
The orientation in $S$ remains unaffected. We also see a significant 
increase in the magnitude of $\delta \mathbf{m}$ in all layers for 
high biases, indicating greater growth in the spin accumulation. 
In the bottom panels we see a behavior in the average spin accumulation in $S$ 
similar to that in the bottom panels of Fig.~\ref{figure5}, with increases to 
the $x$ component for angles $\phi=30^{\circ}$, $90^{\circ}$, 
and $120^{\circ}$. The behavior in $N$ is significantly different 
from that found in the previous cases, where in the $x$ component we 
now see 
 no major critical bias behavior and a steadily decreasing bias dependence: 
this is now similar to the behavior of the magnitude
of the $z$ component. 
The $z$ component has the usual  steady increase with bias, but 
the angular dependence is now most similar to that in 
Fig.~\ref{Fig4c}.  We see then that the angular dependence 
is very sensitive to both the layer thickness and the barriers.

For Fig.~\ref{figure7} we revert to the full set of plots used
e.g. in Fig.~\ref{figure2}, with the same internal organization.
In Fig.~\ref{Fig7a} we see (when comparing with the
results shown in Fig.~\ref{Fig4a} which, as mentioned,
are quite similar to those for the case shown in
Fig.~\ref{figure5}) that when decreasing 
the intermediate ferromagnetic layer spacing, the $x$ and $z$ components 
of the spin current decrease quite significantly in the low bias limit, 
but on the other hand, they increase somewhat in the high bias limit, 
especially the $S_x$ component. The orientation of $\mathbf{S}$ in the superconductor is now rotated closer to the negative $z$ direction, 
much more significantly so for orientations with $\phi > 90^{\circ}$. 
This feature is complemented by Fig.~\ref{Fig7d}, where the average spin 
torque is seen to
increase its rate of growth. This may seem counter-intuitive at first, 
but it is important to note that the 
superconducting pair amplitudes are damped by the ferromagnetic layer. 

In Fig.~\ref{Fig7b} we see, comparing
now directly with Fig.~\ref{figure5}, that decreasing $D_{F2}$ changes the spin 
accumulation in $N$ from a three-peak 
 to a two-peak structure with the same angular dependence and 
greater magnitude. The peaks also  show a greater rotation in orientation compared to those in Fig.~\ref{figure5}, where the spin 
accumulation is more closely aligned to the orientation of ${\bf h_2}$ 
than before. The troughs of these oscillations are still oriented along the $z$ axis.
The overall magnitude of the spin accumulation also increases 
dramatically with bias, at a much greater rate than those in the 
systems discussed previously, as can be seen in Fig.~\ref{Fig7c}. 
However, $\langle\delta m_x\rangle$ in $N$ steadily increases with  bias, with a 
slight peak near the critical bias. The average spin accumulation at 
angle $\phi=150^{\circ}$ does not increase with bias, and remains an outlier.

\section{Conclusions} 
\label{conclusions} 

We have investigated  spin transport for $F/N/F/S$ superconducting 
spin valves. Through our study, we have predicted the main characteristics 
of the relevant spintronic quantities, namely
 the spin current, the spin transfer torque, and the 
local magnetization (a proxy for spin accumulation). We have
done so for multiple variations of the geometrical
and interfacial
 parameters of the spin valve. Our focus has been on samples 
of such thicknesses as can be realistically fabricated,  and which include
 a normal metal spacer and good but imperfect interfaces. The material
parameters employed, such as internal
field and coherence length,  have been shown to be valid
for samples where Nb is the superconductor, 
Cu the normal spacer, and Co the ferromagnet: such
values were successfully used previously to quantitatively 
fit, using our theoretical
methods, the transition temperatures\cite{alejandro} of similar spin 
valve heterostructures.
This quantitative success makes us confident as to the
validity of the predictions presented here.
Our main results are given as a function of position within the spin 
valve, and of the applied bias. We consider
 both  low-bias  values and  the high bias limit where the bias
exceeds the bulk superconductor gap.
We emphasize the dependence of all results
on the misalignment magnetization angle $\phi$ between the $F$ layers;
the misalignment determines the triplet pair formation,  hence 
the range of the proximity effects  and indeed the
valve action. 
 Our analysis includes  variation of the interfacial scattering 
parameters and intermediate layer thicknesses to better encompass 
a full picture of possible real world results. 
However, the parameter space is exceedingly large with no possible extrapolation due to the oscillatory
behavior of many quantities and
the complexity of the self consistent
calculations required. Therefore, what we present here is merely a subset 
of our results with the expressed purpose of establishing the main characteristics 
of the outcomes and exhibiting a glimpse of the richness and
variety of what can  be done.

 Our results are presented in detail in Sec.~\ref{results}.
 We begin by discussing the the dependence of the results
on the scattering potential barriers that would be prevalent in even 
the most ideal fabrication processes. Then, starting
with a  realistic geometry, we vary the intermediate layer 
thicknesses while keeping them within an experimentally realistic range. 
In our results we see a distinct critical bias behavior where, for a certain value of the bias, which is in general $\phi$ dependent and always smaller
than the bulk $S$ gap value, the spin transport
behavior changes, with both the spin current and the spin 
accumulation beginning to penetrate into the superconductor.  
By analyzing the spatially averaged
 spin accumulation and STT within each layer, 
we also see the critical bias behavior featured in the magnitude 
of these quantities. We are then able
to analyze the trends both above and below the
critical bias. These averages show distinct growth in 
the spin accumulation in $S$, and  also in $N$ for certain 
sets of both interfacial scattering and thickness parameters. 
The spin transfer torque also shares this behavior within the ferromagnetic regions, 
with an additional symmetrical behavior in the angular dependence when the interfacial barriers are fully introduced. 

We also observe, at fixed higher bias, 
the spatial precession of the spin current within the 
ferromagnets 
due to the spin transfer torque. The spin current precesses about 
the internal field of the ferromagnet, with a decaying amplitude within the intermediate $F_2$ layer due to the proximity effect of the 
superconductor. This results in both the spin current and the spin accumulation being oriented within the superconductor at 
an angle near the field misalignment angle $\phi$, and at an angle between zero and $\phi$ within the normal metal layer. 
This is only one way in which the misalignment angle plays a factor. Indeed, the critical bias features are  angularly dependent chiefly
because of the angular dependence of the triplet
amplitudes, resulting in  a very
complex and in general non-monotonic behavior in $\phi$ for
 all of our spin transport quantities. The angular dependence of the 
critical bias was already exhibited  in our previous 
results\cite{Moen2017} for the charge current, and they 
correlate with the critical bias features found in the averages. 

Another noteworthy feature of the spin accumulation 
occurs within the normal 
metal layer, where the system transitions, as parameters vary,  from 
a situation where the magnitude of this quantity
has a single peak
at the center of the normal layer, to multiple peak behavior. We 
find that by varying \textit{either} the interfacial scattering 
parameters \textit{or} the normal metal layer thickness, we get a 
transition into a three-peak behavior. Naively, one would assume 
this to be due to the to the normal quantum mechanical effects of the spacial oscillations alone. 
However, by varying the thickness
of the intermediate ferromagnetic layer $D_{F2}$, we see a two-peak 
behavior for the same normal metal layer thickness and interfacial scattering values. This is unique to these spin valve systems, 
which are highly sensitive to the exact set of parameters, both geometrical and physical. Indeed, the spatial spin current and spin accumulation
 features can not be extrapolated to trends within the set of parameters 
we have analyzed. However, the average quantities of the spin accumulation and spin transfer torque may be at least sometimes
extrapolated at high bias values, as the spatial averages tend to be 
quasilinear in this limit.

To conclude, we have calculated both the spin current and spin 
accumulation in superconducting spin valves for a set of experimentally relevant parameters. 
The dependencies of these quantities on the parameters 
(including the misalignment angle $\phi$) are complex, non-monotonic, 
and extremely rich in features. Many of these features are not yet 
fully understood, and only the most prominent ones
have been thoroughly discussed in this work. We expect these results to be a footstool 
onto which more understanding can be developed of the spin transport properties of these nanoscale superconducting spin valves, both through experiment and through continued theoretical work.

\acknowledgments  The authors thank I. Krivorotov (Irvine) and Chien-Te Wu
(National Chiao Tung University) for many helpful 
discussions. This work was supported in part by the US Department of
Energy grant DE-SC0014467.  

\end{document}